\documentclass[
aps,
pra,
twocolumn,
superscriptaddress,
floatfix,
longbibliography
]{revtex4-2}
\usepackage[dvipdfmx]{graphicx}
\usepackage{dcolumn}
\usepackage{bm}
\usepackage{ulem}
\usepackage{amsmath}
\usepackage{amssymb}
\usepackage{txfonts}
\usepackage{hyperref}
\usepackage{color} 
\usepackage{xcolor}
\newcommand{\bs}   {\boldsymbol}

\newcommand{\e}{{\rm e}}
\newcommand{\imag}{{\rm i}}
\newcommand{\dd}{{\rm d}}

\hypersetup{
  colorlinks=true,
  linkcolor=[rgb]{0.60,0.00,0.0},
  citecolor=[rgb]{0.0,0.0,0.6},
  urlcolor=[rgb]{0.0,0.0,0.6},
  setpagesize=false
}

\begin{document}

\title{
  Emergence of a thermal equilibrium in a subsystem of a pure ground state\\
  by quantum entanglement
}

\author{Kazuhiro~Seki}
\affiliation{Computational Quantum Matter Research Team, RIKEN, Center for Emergent Matter Science (CEMS), Saitama 351-0198, Japan}

\author{Seiji~Yunoki}
\affiliation{Computational Quantum Matter Research Team, RIKEN, Center for Emergent Matter Science (CEMS), Saitama 351-0198, Japan}
\affiliation{Computational Materials Science Research Team, RIKEN Center for Computational Science (R-CCS),  Hyogo 650-0047,  Japan}
\affiliation{Computational Condensed Matter Physics Laboratory, RIKEN Cluster for Pioneering Research (CPR), Saitama 351-0198, Japan}

\begin{abstract}
  By numerically exact calculations of spin-1/2 antiferromagnetic
  Heisenberg models on small clusters up to 24 sites, 
  we demonstrate
  that quantum entanglement between subsystems $A$ and $B$ in a pure ground state
  of a whole system $A+B$ can induce thermal equilibrium in subsystem $A$.
  Here, the whole system is bipartitoned with
  the entanglement cut that covers the entire volume of subsystem $A$.
  An effective temperature ${\cal T}_{A}$ of subsystem $A$
  induced by quantum entanglement is {\it not} a parameter but
  can be determined from the entanglement von Neumann entropy $\mathcal{S}_{A}$ 
  and the total energy $\mathcal{E}_{A}$ of subsystem $A$ calculated for the ground state
  of the whole system.  
  We show that temperature
  ${\cal T}_{A}$ can be 
  derived by minimizing the relative entropy for
  the reduced density matrix operator of subsystem $A$ and
  the Gibbs state (i.e., thermodynamic density matrix operator) of subsystem $A$
  with respect to the coupling strength between subsystems $A$ and $B$.
  Temperature ${\cal T}_{A}$
  is essentially identical to the thermodynamic temperature, for which the entropy and 
  the internal energy evaluated using the canonical ensemble in statistical mechanics
  for the isolated subsystem $A$ agree numerically with  
  the entanglement entropy $\mathcal{S}_{A}$ and
  the total energy $\mathcal{E}_{A}$ of subsystem $A$. 
  Fidelity calculations
  ascertain that
  the reduced density matrix operator of subsystem $A$ for
  the pure but entangled ground state of the whole system $A+B$  
  matches, within a maximally $1.5\%$ error in the finite size clusters studied,
  the thermodynamic density matrix operator of subsystem $A$ at temperature ${\cal T}_{A}$,
  despite that these density-matrix operators 
  are different in general.
  We also find that temperature ${\cal T}_{A}$ evaluated
  from the ground state of the whole system 
  depends insignificantly on the system sizes,
  which is consistent with the fact that
  the thermodynamic temperature is an intensive quantity.
  We argue that 
  quantum fluctuation in an entangled pure state can mimic thermal fluctuation in a subsystem. 
  We also provide two simple but nontrivial
  analytical examples of free bosons and free fermions 
  for which the two density-matrix operators are exactly the same 
  if the effective temperature ${\cal T}_{A}$ is adopted. 
  We furthermore discuss implications and possible applications of our finding. 
\end{abstract}

\date{\today}

\maketitle

\section{Introduction}
How thermal equilibrium arises in a pure quantum state has been an attractive 
subject of study in statistical mechanics~\cite{vonNeumann2010}. 
This is often addressed by examining how the time average of an expectation value of observable for 
a pure quantum state after relaxation dynamics 
approaches an ensemble average of the corresponding observable~\cite{Jensen1985,Tasaki1998,Kollar2008,Linden2009,Short2011,Ikeda2015,Pappalardi2017}.
Recently, the eigenstate-thermalization hypothesis
(ETH)~\cite{Deutsch1991,Srednicki1994,Rigol2008,Nandkishore2015,Iyoda2017} 
is widely exploited as a useful concept for investigating 
the thermalization in isolated quantum systems.
The ETH
hypothesizes that expectation values of few-body observables
with respect to energy eigenstates in a given energy shell behave
as microcanonical expectation values of the corresponding energy shell  
(see Ref.~\cite{Deutsch2018} for details). 
However, not all quantum states satisfy the ETH~\cite{Cassidy2011} and 
systems that 
do not follow the ETH can be systematically 
constructed~\cite{Shiraishi2017,shibata2019onsagers}.

The typicality~\cite{Popescu2006,Goldstein2006,Sugita2007,Reimann2007},
which characterizes thermal equilibrium rather than thermalization, 
is also considered as an important concept
for foundation of statistical mechanics. 
The typicality states that 
for almost every pure state randomly sampled from the Hilbert space, 
a single measurement of observable converges 
to the corresponding statistical expectation value with
probability close to $1$ (see Ref.~\cite{Tasaki2016} for detail). 
Based on the typicality, it has been shown that 
statistical mechanics can be formulated in terms of
the thermal pure quantum (TPQ) states~\cite{Sugiura2012,Sugiura2013,Hyuga2014},
rather than conventional mixed states. 
Note that construction of a TPQ state involves multiplications of 
Hamiltonian in non-unitary forms.

Another key ingredient for foundation of statistical mechanics 
from a quantum-mechanical point of view
is the entanglement~\cite{Popescu2006}.
Consider a normalized pure state $|\Psi\rangle$ in
a Hilbert space $\mathcal{H}$, and 
divide the Hilbert space into two, 
$\mathcal{H} = \mathcal{H}_{A} \otimes \mathcal{H}_B$. 
The reduced density matrix operator $\hat{\rho}_{A}$ on $\mathcal{H}_{A}$ 
is defined as $\hat{\rho}_{A} = {\rm Tr}_{B}[|\Psi \rangle \langle \Psi|]$,
where ${\rm Tr}_{B}[\cdot]$ denotes the trace over $\mathcal{H}_{B}$. 
Since $\hat{\rho}_{A}$ is Hermitian 
$(\hat{\rho}_{A}^\dag =\hat{\rho}_{A})$, 
positive semidefinite $(\hat{\rho}_{A} \geqslant 0)$, and
normalized  $({\rm Tr}_{A}[\hat{\rho}_{A}]=1)$, 
it has the form of $\hat{\rho}_{A}=\exp(-\hat{\mathcal{I}}_A)$
with $\hat{\mathcal{I}}_A$ being a Hermitian operator on $\mathcal{H}_{A}$.
$\hat{\mathcal{I}}_A$ is referred to as entanglement Hamiltonian and 
its spectrum~\cite{Ryu2006} is the entanglement spectrum. 
Following Li and Haldane~\cite{Li2008}, 
one can consider $\hat{\rho}_{A}$ as the Gibbs state of ``Hamiltonian'' $\hat{\mathcal{I}}_A$ at
``temperature'' $T=1$.

The entanglement Hamiltonian or the entanglement spectrum 
has been studied for various quantum states, such as
the quantum Hall state~\cite{Ryu2006,Qi2012},
Tomonaga-Luttinger liquids~\cite{Furukawa2011,Lundren2013},
the ground state of the Heisenberg model~\cite{Poilblanc2010,Lauchli2012}, 
the ground state of the Hubbard model~\cite{Toldin2018}, 
the ground states in different phases of magnetic impurity models~\cite{Bayat2014,Shirakawa2016}, 
the valence-bond solid states~\cite{Lou2011}, and
a time-evolved random-product states of the quantum Ising model~\cite{Chang2019}, 
either by numerical or analytical techniques. 
Remarkably, it has been shown for the quantum Hall state
that the entanglement Hamiltonian is proportional to
the Hamiltonian at the boundary~\cite{Qi2012}. 
Also, near the limit of maximal entanglement under certain conditions, 
a proportionality between the entanglement Hamiltonian and the Hamiltonian of
a subsystem has been found~\cite{Peschel2011}.
Moreover, for a wide class of spin models in one-dimensional (1D)
and two-dimensional (2D) lattices,  
it has been shown that an entanglement temperature,
which is defined by means of a field-theoretical approach, 
varies spatially and decreases inversely proportional
to the distance from the entanglement cut
that divides the system into two half spaces
~\cite{Dalmonte2018,Giudici2018,MendesSantos2020}.
These results imply a possibility to find a physical interpretation
for the entanglement Hamiltonian, at least, in some cases. 
Moreover, a recent cold-atom experiment~\cite{Kaufman2016} has shown that
through a unitary evolution of a pure state, 
thermalization occurs on a local scale, and has 
pointed out the importance of the entanglement entropy 
for thermalization.

Such formal similarities between
a reduced density matrix operator and a Gibbs state
may naturally raise a question as to whether 
a thermal equilibrium state in statistical mechanics can emerge 
from a pure quantum state described by quantum mechanics. 
To this end, disentangling the ``temperature''
from the entanglement Hamiltonian in a reduced density matrix operator is a crucial step.  
In this paper, we address this issue 
by numerically analyzing the ground states of spin-1/2 antiferromagnetic 
Heisenberg models in two coupled 1D chains (i.e., two-leg ladder) and in two coupled 
2D square and triangular lattices (i.e., bilayer lattice)
(see Fig.~\ref{ssc} and Fig.~\ref{clusters}).
Under a bipartitioning of the whole system
into subsystems with an entanglement cut that covers 
the entire volume of the subsystem,
our numerical calculations strongly support 
that a thermal equilibrium can emerge
in a partitioned subsystem of a pure ground state with the temperature 
that is not a parameter but is determined by the entanglement von Neumann entropy and the total energy of 
the subsystem. This is further ascertained numerically by the fidelity calculation of the reduced density matrix 
operator and the Gibbs state. 
We also provide two simple but nontrivial examples, relevant to the Unruh effect or 
a two-mode squeezed state in quantum optics and a BCS-type superconducting state, to support this statement analytically.

The rest of the paper is organized as follows. 
In Sec.~\ref{sec.Model}, we introduce the Heisenberg Hamiltonian and 
describe the setup of bipartitioning the system. 
We also briefly review the reduced density matrix operator for a subsystem of a ground state 
and the Gibbs state in the canonical ensemble.  
In Sec.~\ref{sec.results}, we show the numerically exact results revealing that 
the entanglement von Neumann entropy and the total energy of the subsystem are almost identical with
the thermodynamic entropy and the internal energy of the isolated subsystem, respectively, 
provided that a certain form of the effective temperature is introduced. 
Moreover, the fidelity between the reduced density matrix operator and 
the Gibbs state is examined. 
In Sec.~\ref{sec:analytical}, we consider two examples that can be
solved analytically, for which the reduced density matrix operator
is exactly the same as the Gibbs state, 
thus supporting the numerical finding. 
In Sec.~\ref{sec.discussions}, we further discuss the implication of
the
emergent thermal equilibrium in a partitioned subsystem of 
a pure ground state. 
In Sec.~\ref{sec.conclusions}, we conclude the paper with remarks on 
possible application and extension of the present finding. 
Additional discussions on
the effective temperature, the relative entropy, and the thermofield-double state
are given in Appendices~\ref{app}, ~\ref{app.TB}, and ~\ref{app.TFD}, respectively.
Throughout the paper, we set $\hbar=1$ and $k_{B}=1$.

\section{Model and Formalism}\label{sec.Model}
\subsection{Model and bipartitioning}
\begin{figure}
  \begin{center}
    \includegraphics[width=0.75\columnwidth]{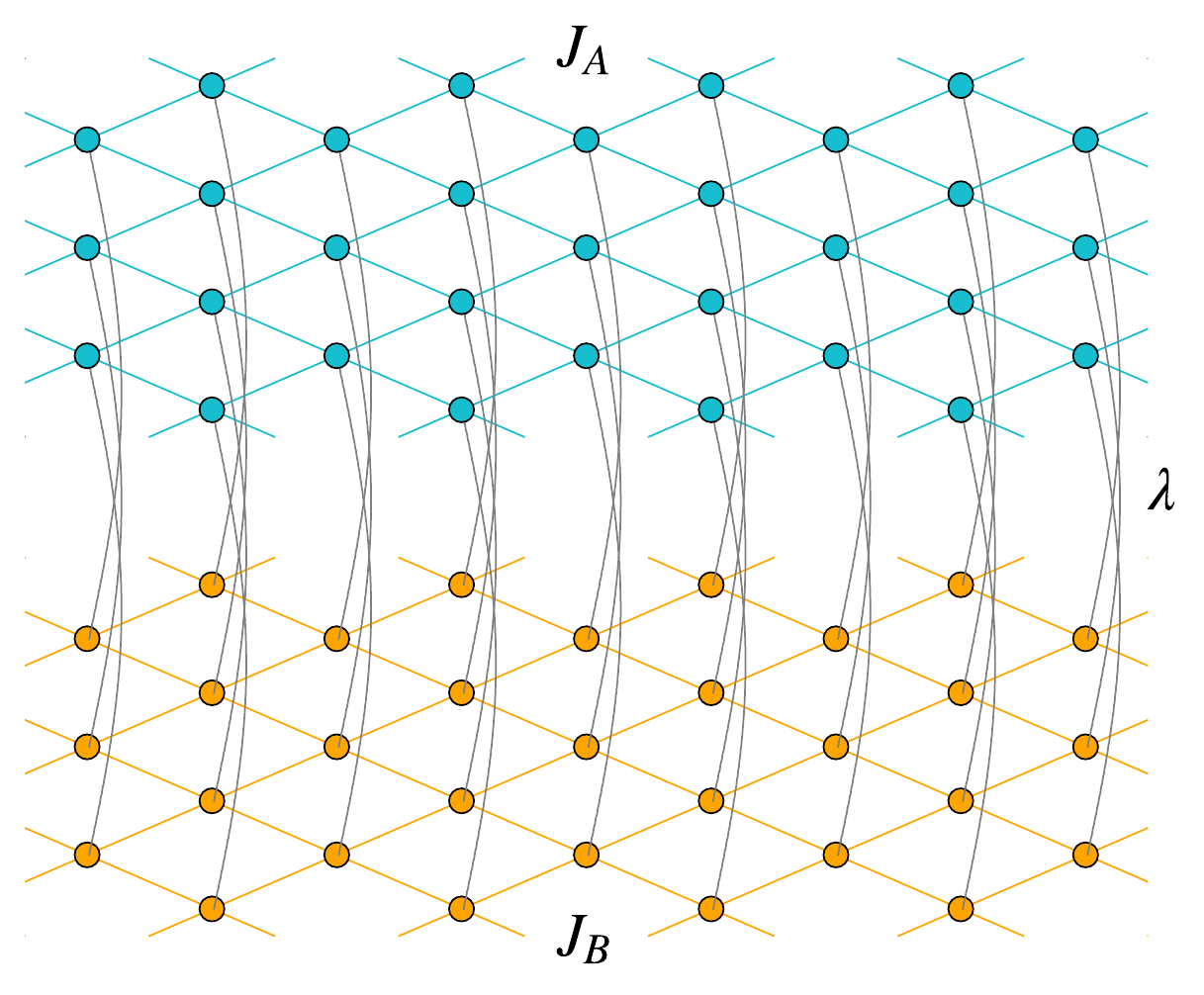}
    \caption{
      Schematic figure of the whole system $A+B$ consisting of
      subsystems $A$ (upper layer indicated by cyan) and $B$ (lower layer indicated by orange),
      which are coupled via the exchange interaction $\lambda$, indicated by 
      curved black lines. The exchange interactions between spins in subsystems $A$ and $B$ are 
      $J_A$ and $J_B$, respectively. 
      In this study, we consider the bipartitioning of the whole system into subsystems
      with the entanglement cut that covers the entire volume of subsystem $A$. 
      \label{ssc}}    
  \end{center}
\end{figure}

We consider the spin-1/2 antiferromagnetic Heisenberg model
described by the following Hamiltonian: 
\begin{eqnarray}
  \hat{H} &=&
  \sum_{\langle i,j \rangle} J_{ij} \ 
  \hat{\bs{S}}_{i} \cdot \hat{\bs{S}}_{j}
\end{eqnarray}
where
$\langle i,j \rangle$
runs over all pairs of nearest-neighbor sites $i$ and $j$ 
in two coupled 1D chains (i.e., two-leg ladder) or in two coupled 
2D square or triangular lattices (i.e., bilayer lattice). 
$\bs{\hat{S}}_{i}$ is the spin-$1/2$ operator located at the $i$th 
site and the nearest-neighbor spins are connected with 
the exchange interaction 
$J_{ij}=J_{A}$, $J_{B}$, or $\lambda$ (see Fig.~\ref{ssc}).
We denote by $N$ the number of spins 
and thus the dimension of the total Hilbert space $\mathcal{H}$ is $D=\dim \mathcal{H}=2^N$. 
We consider the case where the exchange interactions are antiferromagnetic ($J_{ij}>0$).

To study the entanglement in the ground state of $\hat{H}$, 
we bipartition the Hilbert space $\mathcal{H}$ of the whole system
into those of subsystems $A$ and $B$ as 
$\mathcal{H}=\mathcal{H}_{A} \otimes \mathcal{H}_{B}$. 
Accordingly, the Hamiltonian $\hat{H}$ can be written as 
\begin{equation}
  \hat{H}(\lambda)
  = \hat{H}_{A} \otimes \hat{I}_B
  + \hat{I}_A \otimes \hat{H}_{B}
  + \hat{V}_{AB}(\lambda),
  \label{fullHam}
\end{equation}
where 
\begin{alignat}{1}
  &\hat{H}_{A} = J_{A} \sum_{\langle i,j\rangle \in A} \hat{\bs{S}}_{i} \cdot \hat{\bs{S}}_{j}, \\
  &\hat{H}_{B} = J_{B} \sum_{\langle i,j\rangle \in B} \hat{\bs{S}}_{i} \cdot \hat{\bs{S}}_{j}, \\
  &\hat{V}_{AB}(\lambda) = \lambda \sum_{\langle i,j \rangle, i\in A, j \in B} \hat{\bs{S}}_{i} \cdot \hat{\bs{S}}_{j}, 
\end{alignat}
and 
$\hat{I}_{A\,(B)}$ is the identity operator on $\mathcal{H}_{A\,(B)}$ (see Fig.~\ref{ssc}). 
$\hat{H}_{A\,(B)}$ is the Hamiltonian of subsystem $A$~$(B)$ and 
$\hat{V}_{AB}(\lambda)$ describes the exchange interaction 
between subsystems $A$ and $B$.
The subsystem $B$ considered here is essentially a copy of the subsystem $A$ except 
that its interaction strength $J_B$ may differ from $J_A$.
We denote
by $N_{A\,(B)}$ the number of spins in subsystem $A$~$(B)$ and 
thus the dimension of the Hilbert space for subsystem $A$~$(B)$ is 
$D_{A\,(B)}=\dim \mathcal{H}_{A\,(B)}=2^{N_{A\,(B)}}$. 
Note that $N=N_{A}+N_{B}$ and $D=D_A D_B$. 
The exchange interaction $\lambda$ controls 
the entanglement between subsystems $A$ and $B$.
As shown schematically in Fig.~\ref{ssc},
the whole system is bipartitioned into two subsystems
$A$ and $B$ with the entanglement cut that 
covers the entire volume of subsystem $A$.

Let $|\Psi_0 (\lambda)\rangle$ be the normalized ground state
of $\hat{H}(\lambda)$.
Note that the $\lambda$ dependency of $\hat{H}$  and $|\Psi_0 \rangle$ 
is explicitly denoted 
since we consider the entanglement 
between subsystems $A$ and $B$ in the ground state $|\Psi_0(\lambda)\rangle$
with varying $\lambda$. 
Although the ground state should depend on the exchange interactions as 
$|\Psi_0 \rangle = |\Psi_0(J_B/J_A,\lambda/J_A) \rangle$,  
here we simply assume the $J_A$ and $J_B$ dependence of these quantities.

Four remarks are in order. In our setup, 
(i)
we do not assume any finite temperature in
the subsystem $A$ or $B$,
(ii) the volume $N_{B}$ of subsystem $B$ is not necessarily
sufficiently larger than the volume $N_{A}$ of subsystem $A$ (and vice versa), 
(iii)
the coupling term $\hat{V}_{AB}(\lambda)$
between subsystems $A$ and $B$ is not necessarily small 
as compared to $\hat{H}_A$ and $\hat{H}_B$,
and 
(iv) a pure state of the whole system $A+B$ is always chosen as
its ground state $|\Psi_0(\lambda)\rangle$,
and thus any stochastic sampling of pure states from $\mathcal{H}$ does not apply. 
Remarks (i)-(iii) imply that the role of subsystem $B$ 
is not the heat bath for subsystem $A$, unlike in the conventional statistical mechanics. 
Remark (iv) implies that our approach do not make use of 
the typicality argument.

\subsection{Entanglement entropy and energy of subsystem $A$}\label{sec.ent}
The ground state can be expanded as 
\begin{equation}
  |\Psi_0 (\lambda)\rangle
  = \sum_{i=1}^{D} c_i(\lambda)|i \rangle
  = \sum_{i=1}^{D_A} \sum_{j=1}^{D_B} c_{i,j}(\lambda)
  |{i}\rangle_A
  |{j}\rangle_B, 
\end{equation}
where
$\{|i\rangle\}_{i=1}^{D}$ is the orthonormal basis set in $\mathcal{H}$, and 
$\{|i\rangle_{A\,(B)}\}_{i=1}^{D_{A\,(B)}}$ is the orthonormal basis set in $\mathcal{H}_{A\,(B)}$. 
The coefficients $\{c_i(\lambda)\}_{i=1}^{D}$ are rewritten 
as $\{c_{i,j}(\lambda)\}_{i=1,j=1}^{D_A,D_B}$
simply by using the labels 
$i$ and $j$ for subsystems $A$ and $B$.  
The reduced density matrix  operator $ \hat{\rho}_A^{\rm red}$ of
subsystem $A$ is now given as 
\begin{alignat}{1}
  \hat{\rho}_A^{\rm red}(\lambda)
  &=
      {\rm  Tr}_B\left[
        |\Psi_0(\lambda) \rangle \langle \Psi_0(\lambda) |\right] \notag \\
      &=
      \sum_{k=1}^{D_B} {_B}\langle k|\Psi_0(\lambda) \rangle \langle \Psi_0(\lambda)
      | k \rangle_B \notag \\
      &=
      \sum_{i=1}^{D_A} \sum_{j=1}^{D_A}
      \rho^{\rm red}_{A,ij}(\lambda)
          {|i\rangle_A} {_A}\langle j|, 
\end{alignat}
where the reduced density matrix
\begin{equation} 
  \rho^{\rm red}_{A,ij}(\lambda)
  \equiv \sum_{k=1}^{D_B} c_{i,k}(\lambda) c_{j,k}^*(\lambda)
  \label{rhoA}
\end{equation}
is introduced. 
With a $D_A \times D_B$ matrix $\bs{c}(\lambda)$ defined 
as $[\bs{c}(\lambda)]_{ij}=c_{i,j}(\lambda)$,
the reduced density matrix can be written in a matrix form as 
$\bs{\rho}^{\rm red}_{A}(\lambda) = \bs{c}(\lambda) \bs{c}(\lambda)^\dag$.
The reduced density matrix
$\bs{\rho}^{\rm red}_A(\lambda)$
is Hermitian and positive semidefinite, and 
satisfies ${\rm Tr}[\bs{\rho}^{\rm red}_A(\lambda)]
=\langle \Psi_0 (\lambda)| \Psi_0(\lambda) \rangle=1$~\cite{Fano1957}. 
The positive semidefiniteness of
$\bs{\rho}^{\rm red}_A(\lambda)$ follows from
the fact that $\bs{\rho}^{\rm red}_A(\lambda)$
is a Gram matrix as apparently noticed in 
Eq.~(\ref{rhoA}).

The entanglement entropy $\mathcal{S}_{A}(\lambda)$
of subsystem $A$ is here defined as the von Neumann entropy
of the reduced density matrix operator,  
\begin{equation}
  \mathcal{S}_{A}(\lambda) 
  \equiv
      {\rm Tr}_{A} \left[\hat{\rho}^{\rm red}_{A}(\lambda)
        \hat{\mathcal{I}}^{\rm red}_{A}(\lambda)\right]
      =
      -{\rm Tr}_{A}\left[ \hat{\rho}^{\rm red}_A(\lambda) \ln { \hat{\rho}^{\rm red}_A(\lambda)} \right]       
\end{equation}
with $\hat{\mathcal{I}}_{A}^{\rm red}(\lambda)=-\ln \hat{\rho}^{\rm red}_{A}(\lambda)$
being the entanglement Hamiltonian.
The entanglement entropy satisfies $0 \leqslant \mathcal{S}_{A} \leqslant \ln D_{A}$, where 
the lower bound is achieved when $\hat{\rho}_{A}^{\rm red}$ is a pure state and
the upper bound is obtained when $\hat{\rho}_{A}^{\rm red}$ is the maximally mixed state. 
The energy $\mathcal{E}_A(\lambda)$
of subsystem $A$ is calculated as 
\begin{equation}
  \mathcal{E}_A(\lambda) \equiv
          {\rm Tr}_{A}\left[\hat{\rho}^{\rm red}_{A}(\lambda) \hat{H}_{A}
            \right]
          =
          \langle \Psi_0(\lambda) | \hat{H}_{A}\otimes \hat{I}_B |\Psi_0(\lambda) \rangle. 
\end{equation}
Note that these quantities are defined using 
the ground-state wavefunction $|\Psi_0(\lambda) \rangle$
of the whole system $\hat{H}(\lambda)$.

\subsection{Canonical ensemble of subsystem $A$}\label{sec.can}
Let us consider the canonical ensemble in statistical mechanics 
for the isolated subsystem $A$ without subsystem $B$. 
In the canonical ensemble, the heat bath with temperature $T$ is assumed and 
the average of an observable $\hat{\mathcal{O}}$ in subsystem $A$ 
is given as 
\begin{equation}
  \left\langle \hat{\mathcal{O}} \right\rangle_{\beta}^{\rm can} \equiv
          {\rm Tr}_{A}\left[\hat{\rho}_A^{\rm can}(\beta) \hat{\mathcal{O}}\right], 
\end{equation}
where
\begin{equation}
  \hat{\rho}_A^{\rm can}(\beta) \equiv \frac{\e^{-\beta \hat{H}_A} }{Z_A(\beta)}
  \label{rho_can}
\end{equation}
is the Gibbs state, i.e., thermodynamic density matrix operator,   
$\beta=1/T$ is the inverse temperature, and 
$Z_{A}(\beta)={\rm Tr}_{A}\left[ \e^{-\beta \hat{H}_A} \right]$
is the partition function. 
The entropy $S_{A}(\beta)$ and the internal energy $E_{A}(\beta)$ 
are given, respectively, as
\begin{alignat}{1}
  S_{A}(\beta)&=
  \left \langle \mathcal{I}^{\rm can}_A(\beta)
  \right \rangle_{\beta}^{\rm can}
  =-{\rm Tr}_A \left[
    \hat{\rho}_A^{\rm can}(\beta)
    \ln \hat{\rho}_{A}^{\rm can}(\beta) \right]
\end{alignat}
and
\begin{alignat}{1}
  E_{A}(\beta)
  &= \left \langle \hat{H}_{A} \right \rangle_{\beta}^{\rm can}
  = {\rm Tr}_A\left[\hat{\rho}_{A}^{\rm can}(\beta) \hat{H}_{A}  \right] 
\end{alignat}
with $\hat{\mathcal{I}}_A^{\rm can}(\beta)=-\ln \hat{\rho}_{A}^{\rm can}(\beta)$.

\section{Numerical Results}\label{sec.results}
By numerically analyzing the Heisenberg models described above, 
we now 
examine
the emergence of a
thermal equilibrium 
in the partitioned subsystem $A$ by quantum entanglement. 
First, we briefly review two limiting cases when 
$\lambda=0$ (i.e., zero entanglement limit) and 
$\lambda=\infty$ (i.e., maximal entanglement limit). 
Next, we show numerical results 
for general values of $\lambda$, 
which support the emergent thermal equilibrium, 
provided that the temperature is appropriately introduced.

\subsection{Zero entanglement and zero-temperature limit}\label{sec.zeroT}
When $\lambda=0$, there exists no entanglement between subsystems $A$ and $B$ 
and any eigenstate of $\hat{H}(\lambda=0)$ is separable
under the bipartioning of the system considered here.
In particular, the ground state $|\Psi_0 (\lambda=0)\rangle$ is the product state
of the ground states $|\psi_0^A \rangle$ and $|\psi_0^B \rangle$
of subsystems $A$ and $B$, respectively,
i.e., 
$|\Psi_0(\lambda=0)\rangle = |\psi_0^A \rangle |\psi_0^B \rangle$. 
Thus, the subsystem $A$ is a pure state and the reduced density matrix operator of subsystem $A$ is 
\begin{equation}
  \hat{\rho}_A^{\rm red}(\lambda=0) = |\psi^A_0 \rangle \langle \psi^A_0 | 
\end{equation}
with the entanglement von Neumann entry of subsystem $A$   
\begin{equation}
  \mathcal{S}_A(\lambda=0)=0, 
\end{equation}
which is the lower bound of $\mathcal{S}_A$.
The thermodynamic density matrix operator 
in the zero-temperature limit is apparently identical 
with the reduced density matrix operator at $\lambda=0$, i.e., 
\begin{equation}
  \hat{\rho}_{A}^{\rm can}(\beta=\infty) =
  \hat{\rho}_{A}^{\rm red}(\lambda=0),    
\end{equation}
and the thermodynamic entropy is $S_{A}(\beta=\infty)=0$.

\subsection{Maximal entanglement and infinite-temperature limit}\label{sec.infiniteT}
When $\lambda = \infty$,
the total Hamiltonian $\hat{H}(\lambda)$ is dominated by $\hat{V}_{AB}(\lambda)$ and 
the corresponding ground state $|\Psi_0 (\lambda=\infty) \rangle$ 
is a singlet-pair product state, i.e., the direct product of 
the spin singlet states formed by two neighboring spins, each locating in subsystems $A$ and $B$. 
Thus, $|\Psi_0 (\infty) \rangle$ 
has the maximal entanglement between subsystems $A$ and $B$. 
After tracing out subsystem $B$, the subsystem $A$
is described by the maximally mixed state 
(see a similar argument in Ref.~\cite{White2009}) 
and the reduced density matrix operator of subsystem $A$ is  
\begin{equation}
  \hat{\rho}_A^{\rm red}(\lambda=\infty) 
  = \frac{1}{D_A} \hat{I}_A   
\end{equation}
with the entanglement von Neumann entropy of subsystem $A$ 
\begin{equation}
  \mathcal{S}_A(\lambda=\infty)=\ln{D_A}=N_A \ln 2, 
\end{equation}
which is the upper bound of $\mathcal{S}_A$. 
The thermodynamic density matrix operator 
in the infinite-temperature limit is identical 
with the reduced density matrix operator at $\lambda=\infty$, i.e., 
\begin{equation}
  \hat{\rho}_{A}^{\rm can}(\beta=0) =
  \hat{\rho}_{A}^{\rm red}(\lambda=\infty),    
\end{equation}
and the thermodynamic entropy is $S_{A}(\beta=0)=N_A \ln 2$.

\subsection{General values of $\lambda$ and $\beta$}\label{sec.generalT}
\begin{figure}
  \begin{center}
    \includegraphics[width=0.85\columnwidth]{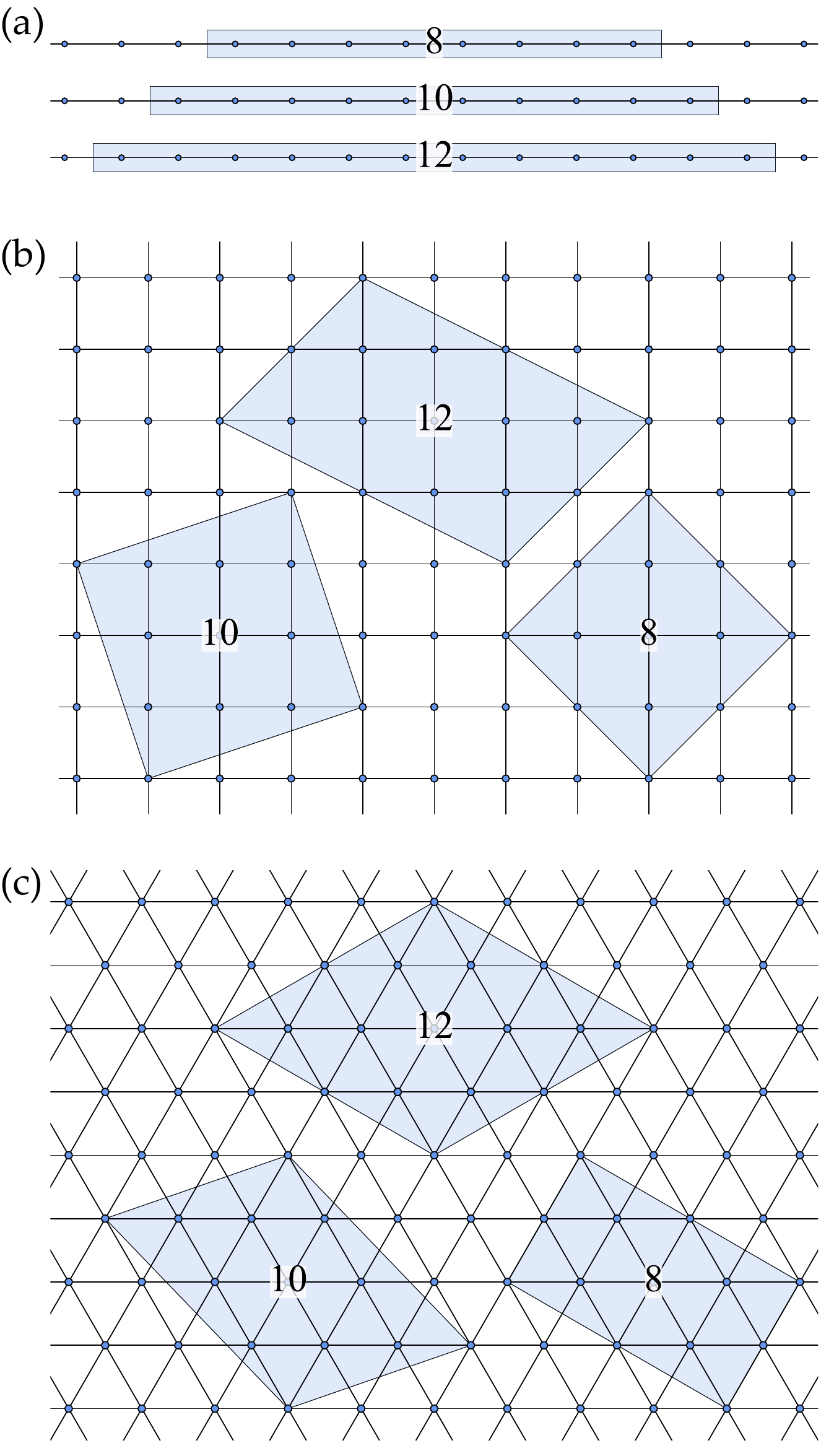}
    \caption{
      Finite-size clusters of $N_A=8, 10,$ and $12$ on 
      (a) the 1D chain,
      (b) the 2D square lattice, and
      (c) the 2D triangular lattice
      used for the calculations.       
      Periodic-boundary conditions are applied for all clusters.
      The figures show only the subsystem $A$, which couples to   
      the subsystem $B$ with $N_B=N_A$ when the ground state of the whole system 
      ${\hat H}(\lambda)={\hat H}_A + {\hat H}_B + {\hat V}_{AB}(\lambda)$ is 
      calculated (see Fig.~\ref{ssc}).
      \label{clusters}}    
  \end{center}
\end{figure}

We calculate the ground state $|\Psi_0 (\lambda)\rangle$ of $\hat{H}(\lambda)$
by the Lanczos method, and evaluate 
$\mathcal{S}_A(\lambda)$ and $\mathcal{E}_{A}(\lambda)$ of subsystem $A$, 
accordingly to the formalism described in Sec.~\ref{sec.ent}.
To examine how these entanglement-induced
quantities of the ground state $|\Psi_0 (\lambda)\rangle$ can be related to 
the thermodynamic quantities of the subsystem,
we also calculate $S_A(\beta)$ and $E_A(\beta)$ of the isolated subsystem $A$ 
as a function of the inverse temperature $\beta=1/T$ by 
numerically diagonalizing the Hamiltonian $\hat{H}_A$ (see Sec.~\ref{sec.can}). 
The finite-size systems used for these calculations are shown in Fig.~\ref{clusters}.
We should emphasize that the ground state $|\Psi_0 (\lambda)\rangle$ of $\hat{H}(\lambda)$ is 
spin singlet (i.e., total spin and thus the $z$ component of the total spin being both zero) and the 
total momentum of $|\Psi_0 (\lambda)\rangle$ is zero, while the canonical ensemble described in Sec.~\ref{sec.can} 
averages over all eigenstates of $\hat{H}_A$ in all spin and momentum symmetry sectors.

\begin{figure*}
  \begin{center}
    \includegraphics[width=1.9\columnwidth]{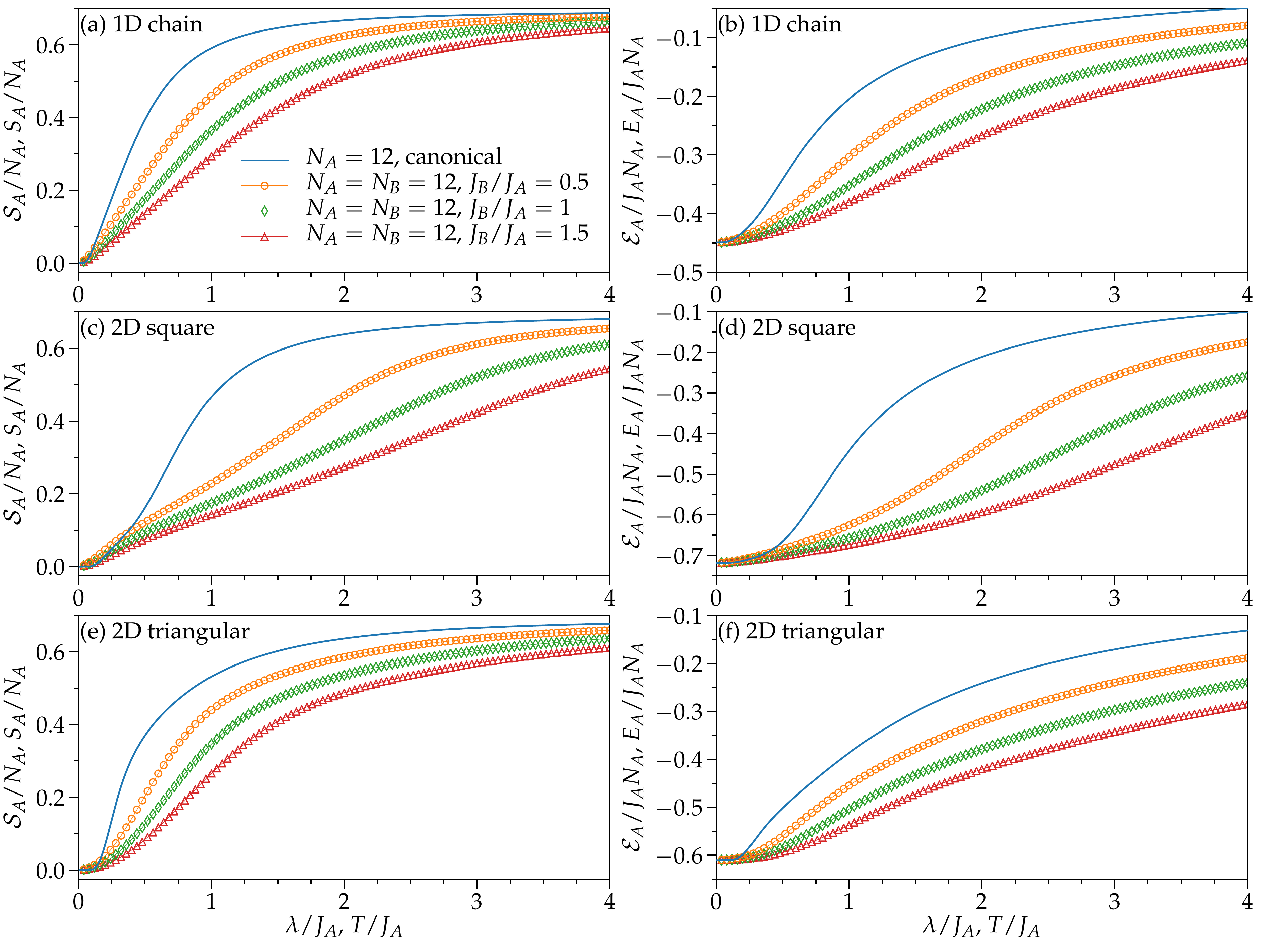}
    \caption{
      (a,c,e)
      The entanglement von Neumann entropy
      per site $\mathcal{S}_{A}(\lambda)/N_A$
      of subsystem $A$ as
      a function of $\lambda/J_{A}$ for several values of $J_B$ (symbols) and
      the thermodynamic entropy
      per site $S_{A}(\beta)/N_A$
      of the isolated subsystem
      $A$ as a function of temperature $T$ (solid line).
      (b,d,f)
      The energy
      per site $\mathcal{E}_{A}(\lambda)/N_A$
      of subsystem $A$ as
      a function of $\lambda/J_{A}$ for several values of $J_B$ (symbols) and
      the internal energy
      per site $E_{A}(\beta)/N_A$ of the isolated subsystem $A$ as a function of temperature $T$ (solid line).
      In (a) and (b), $\mathcal{S}_{A}(\lambda)$ and $\mathcal{E}_{A}(\lambda)$ are calculated for the two coupled 1D 
      chains (i.e., two-leg ladder) 
      with $N_{A}=N_{B}=12$, and $S_{A}(\beta)$ and $E_{A}(\beta)$ are calculated for the 1D chain with $N_{A}=12$. 
      In (c) and (d), $\mathcal{S}_{A}(\lambda)$ and $\mathcal{E}_{A}(\lambda)$ are calculated for the two coupled 2D square 
      lattices (i.e., bilayer square lattice) 
      with $N_{A}=N_{B}=12$, and $S_{A}(\beta)$ and $E_{A}(\beta)$ are calculated for the 2D square lattice with $N_{A}=12$.
      In (e) and (f), $\mathcal{S}_{A}(\lambda)$ and $\mathcal{E}_{A}(\lambda)$ are calculated for the two coupled 2D triangular 
      lattices (i.e., bilayer triangular lattice) 
      with $N_{A}=N_{B}=12$, and $S_{A}(\beta)$ and $E_{A}(\beta)$ are calculated for the 2D triangular lattice with $N_{A}=12$.
      \label{JvsSE}}    
  \end{center}
\end{figure*}

Figure~\ref{JvsSE} shows the $\lambda$ dependence of 
$\mathcal{S}_A(\lambda)$ and $\mathcal{E}_A(\lambda)$ with different values of $J_B/J_A$. 
For comparison, the $T$ dependence of  
$S_A(\beta)$ and $E_A(\beta)$ is also shown. 
It is clearly observed that 
$\mathcal{S}_A(\lambda)$ and $\mathcal{E}_A(\lambda)$ increase monotonically with $\lambda$.
Moreover, as discussed in Sec.~\ref{sec.zeroT} and
Sec.~\ref{sec.infiniteT}, the ranges of
the entanglement von Neumann entropy $\mathcal{S}_{A}(\lambda)$ and the thermodynamic entropy 
$S_A(\beta)$ with varying $\lambda$ and $T$, respectively, 
as well as those of $\mathcal{E}_{A}(\lambda)$ and $E_A(\beta)$, agree with each other.

\begin{figure*}
  \begin{center}
    \includegraphics[width=1.9\columnwidth]{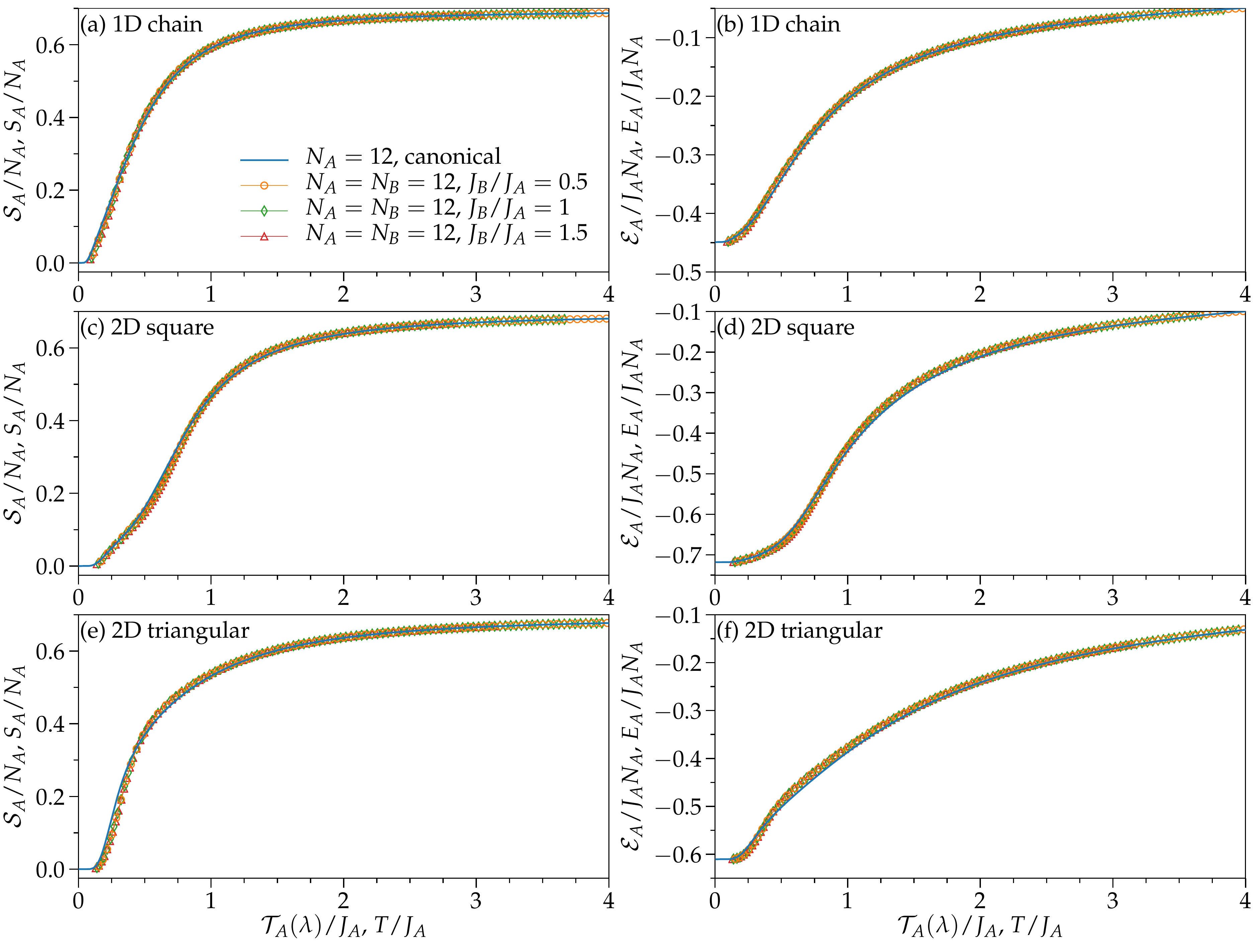}
    \caption{
      Same as Fig.~\ref{JvsSE} but $\mathcal{S}_{A}(\lambda)$ and $\mathcal{E}_{A}(\lambda)$ are now plotted 
      as a function of $\mathcal{T}_A(\lambda)$. 
      \label{SandE}}    
  \end{center}
\end{figure*}

Figure~\ref{SandE} shows the same quantities  
$\mathcal{S}_A(\lambda)$ and $\mathcal{E}_A(\lambda)$ but
as a function of an effective temperature defined as 
$\mathcal{T}_{A}(\lambda)=1/\mathcal{B}_{A}(\lambda)$ with 
\begin{equation}
  \mathcal{B}_{A}(\lambda) =
  \lim_{\Delta \lambda \to 0}
  \frac
      {\mathcal{S}_{A}(\lambda+\Delta \lambda)-\mathcal{S}_{A}(\lambda)}
      {\mathcal{E}_{A}(\lambda+\Delta \lambda)-\mathcal{E}_{A}(\lambda)}
      =\frac
      {\partial_{\lambda} \mathcal{S}_{A}(\lambda)}
      {\partial_{\lambda} \mathcal{E}_{A}(\lambda)}.
  \label{beta_dSdE}
\end{equation}
In the numerical calculations,
we evaluate $\mathcal{B}_{A}(\lambda)$ for $\lambda \geqslant \Delta\lambda$
by a central finite difference 
$\mathcal{B}_{A}(\lambda) \approx
(\mathcal{S}_{A}(\lambda+\Delta \lambda)-\mathcal{S}_{A}(\lambda-\Delta \lambda))/
(\mathcal{E}_{A}(\lambda+\Delta \lambda)-\mathcal{E}_{A}(\lambda-\Delta \lambda))$
with $\Delta \lambda/J_{A}=0.02$. 
For comparison, the $T$ dependence of  
$S_A(\beta)$ and $E_A(\beta)$ of the canonical ensemble
is also shown in Fig.~\ref{SandE}. 
Remarkably, for each lattice structure,
$\mathcal{S}_{A}(\lambda)$ for all $J_B/J_A$ values
are on a universal curve 
[Figs.~\ref{SandE}(a), \ref{SandE}(c), and \ref{SandE}(e)].
Moreover, such a universal curve essentially coincides with
the temperature dependence of the thermodynamic entropy $S_A(\beta)$
for the corresponding lattice. 
The same is also found in
the energy $\mathcal{E}_A(\lambda)$ and
the internal energy $E_A(\beta)$,
as shown in Figs.~\ref{SandE}(b), \ref{SandE}(d), and \ref{SandE}(f).
Note that a lack of data points around the limit of $\mathcal{T}_A(\lambda)=0$ in Fig.~\ref{SandE} is 
due to the finite-difference scheme employed for evaluating $\mathcal{B}_A(\lambda)=\mathcal{T}_A(\lambda)^{-1}$ 
in Eq.~(\ref{beta_dSdE}). If smaller $\Delta \lambda$ is chosen, one may find more data points around this limit.

\begin{figure}
  \begin{center}
    \includegraphics[width=0.95\columnwidth]{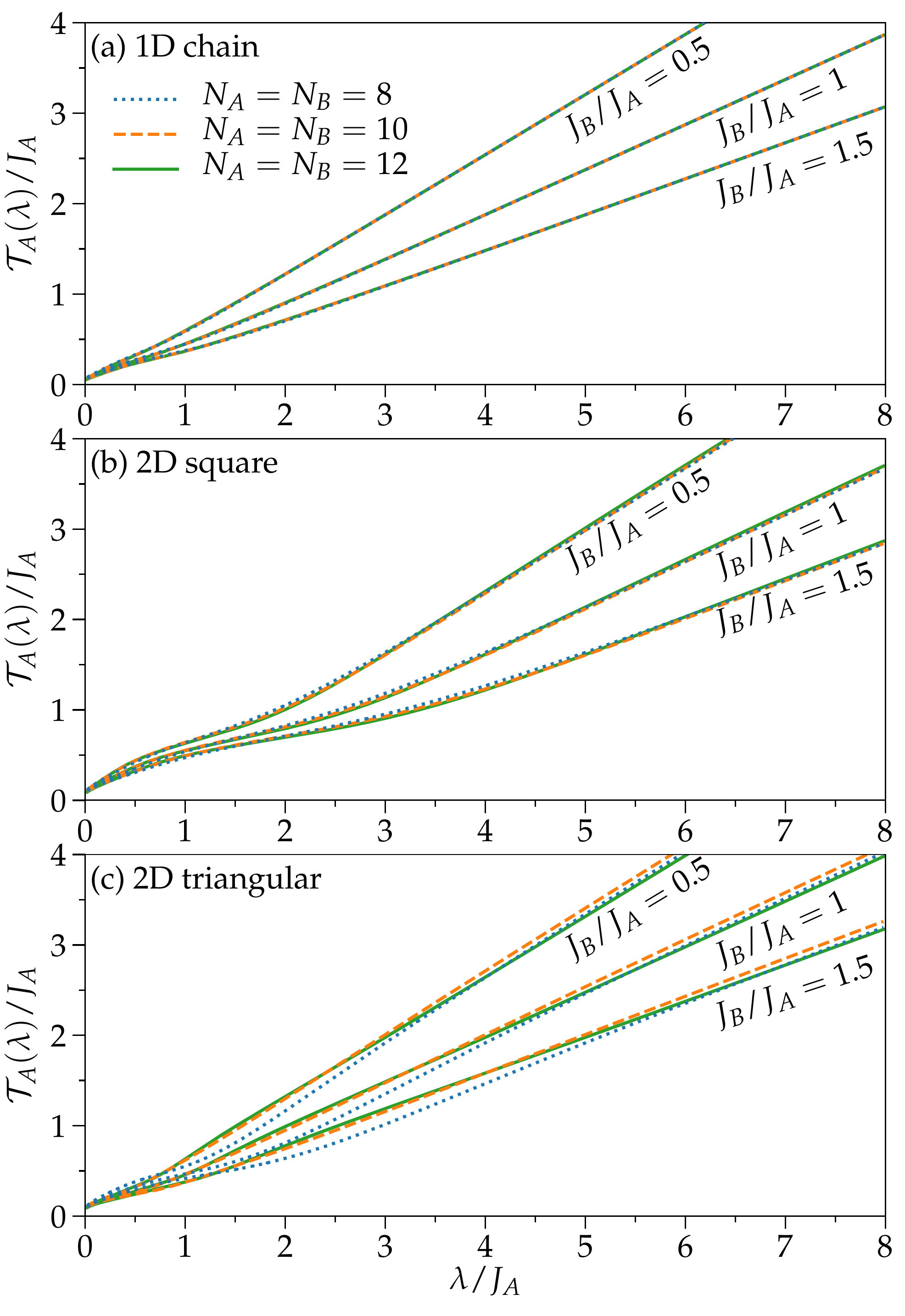}
    \caption{
      The effective temperature $\mathcal{T}_A(\lambda)$
      as a function of $\lambda$ 
      calculated for (a) the two coupled 1D chains (i.e., two-leg ladder), 
      (b) the two coupled 2D square lattices (i.e., bilayer square lattice), and 
      (c) the two coupled 2D triangular lattices (i.e., bilayer triangular lattice) 
      with $N_A=N_B=8, 10,$ and $12$. 
      The results for 
      $J_B/J_A=0.5$, 1, and 1.5 are shown
      (from top to bottom in each panel).
      \label{fig.temperature}}
  \end{center}
\end{figure}

Figure~\ref{fig.temperature} shows 
the effective temperature $\mathcal{T}_A(\lambda)$
as a function of $\lambda$.
While the dependence of $\mathcal{T}_A(\lambda)$
on $\lambda$ is nontrivial, $\mathcal{T}_A(\lambda)$ increases monotonically with $\lambda$. 
We can also notice that 
the increase of 
$\mathcal{T}_A(\lambda)$ is 
more significant for smaller $J_B/J_A$.  
Furthermore, 
we find that $\mathcal{T}_A(\lambda)$
depends insignificantly 
on the system size,
as expected for an analog to the
thermodynamic temperature, which is an intensive quantity. 
A relatively large deviation of 
$\mathcal{T}_A(\lambda)$ for the smallest cluster ($N_A=N_B=8$),  
observed in Fig.~\ref{fig.temperature}(c) for the  
triangular lattice 
at low effective temperatures $\mathcal{T}_A(\lambda)/J_A \lesssim 2$,  
might be due to the
strong finite-size effect.

We note that the form of Eq.~(\ref{beta_dSdE}) for the effective temperature 
is apparently analogous to the definition
of the inverse temperature in thermodynamics~\cite{Lieb1999},  
except that there is the microscopic parameter $\lambda$, through which
$\mathcal{S}_A(\lambda)$ and $\mathcal{E}_A(\lambda)$ are mediated.
Indeed, Eq.~(\ref{beta_dSdE}) can be derived
by minimizing the relative entropy for
the reduced density matrix operator and
the thermodynamic density matrix operator
with respect to $\lambda$. 
Here, the relative entropy for density matrix operators
$\hat{\rho}_0$ and $\hat{\rho}_1$ is defined as  
$D(\hat{\rho}_1|\hat{\rho}_0)= 
{\rm Tr}\left[\hat{\rho}_1 \ln \hat{\rho}_1 \right]
-{\rm Tr}\left[\hat{\rho}_1 \ln \hat{\rho}_0 \right]$ with 
the trace taken over the Hilbert space on which 
$\hat{\rho}_0$ and $\hat{\rho}_1$ are defined 
(also see Appendix~\ref{app}). 
Substituting 
$\hat{\rho}_0 = \hat{\rho}_{A}^{\rm can}(\beta)$ and
$\hat{\rho}_1 = \hat{\rho}_{A}^{\rm red}(\lambda)$ into 
$D(\hat{\rho}_1|\hat{\rho}_0)$, we obtain that 
\begin{equation}
  D[\hat{\rho}_A^{\rm red}(\lambda)|\hat{\rho}_A^{\rm can}(\beta)]
  =-\mathcal{S}_A(\lambda)+\beta \mathcal{E}_A(\lambda) + \ln{Z_A(\beta)}.
  \label{relative}
\end{equation}
It is interesting to notice that there is 
the ``cross term'' in the right hand side of Eq.~(\ref{relative}) 
between quantities representing the quantum entanglement, 
i.e., the ground state energy $\mathcal{E}_A(\lambda)$ of subsystem $A$, 
and thermodynamics, i.e., the inverse temperature $\beta$.  
The minimization of $D[\hat{\rho}_A^{\rm red}(\lambda)|\hat{\rho}_A^{\rm can}(\beta)]$
with respect to $\lambda$, i.e., 
$\frac{\partial D[\hat{\rho}_A^{\rm red}(\lambda)|\hat{\rho}_A^{\rm can}(\beta)]}{\partial \lambda}=0$, 
yields
\begin{equation}
  \beta= \frac
         {\partial_{\lambda} \mathcal{S}_{A}(\lambda)}
         {\partial_{\lambda} \mathcal{E}_{A}(\lambda)} = \mathcal{B}_A(\lambda).
         \label{minimize}
\end{equation}
Therefore, when we consider $\lambda$ as a variable, $\beta$ should be determined as in Eq.~(\ref{beta_dSdE}). 
We can also minimize 
$D[\hat{\rho}_A^{\rm red}(\lambda)|\hat{\rho}_A^{\rm can}(\beta)]$ with respect to $\beta$, which 
yields 
\begin{equation}
  \mathcal{E}_A(\lambda)=E_A(\beta), 
  \label{eq:minimize2}
\end{equation}
suggesting that when we consider $\beta$ as a variable, $\lambda$ should be determined so as to satisfy 
Eq.~(\ref{eq:minimize2}). Although we do not adopt the latter, it turns out that Eq.~(\ref{eq:minimize2}) is almost perfectly 
satisfied in our numerical calculations shown in Fig.~\ref{SandE}, if $\beta$ is determined by Eq.~(\ref{beta_dSdE}).    
It should be noted that the relative entropy is not symmetric, i.e.,
$D(\hat{\rho}_1|\hat{\rho}_0) \not =D(\hat{\rho}_0|\hat{\rho}_1)$ for $\hat{\rho}_0\not=\hat{\rho}_1$, and  
the minimization of 
$D[\hat{\rho}_A^{\rm can}(\beta)|\hat{\rho}_A^{\rm red}(\lambda)]$
with respect to either $\lambda$ or $\beta$ 
does not yield Eq.~(\ref{minimize}). 
Further discussions on $\mathcal{B}_A(\lambda)$ in
the form of Eq.~(\ref{beta_dSdE}) are given in Sec.~\ref{sec:beta}.

\begin{figure}
  \begin{center}
    \includegraphics[width=0.9\columnwidth]{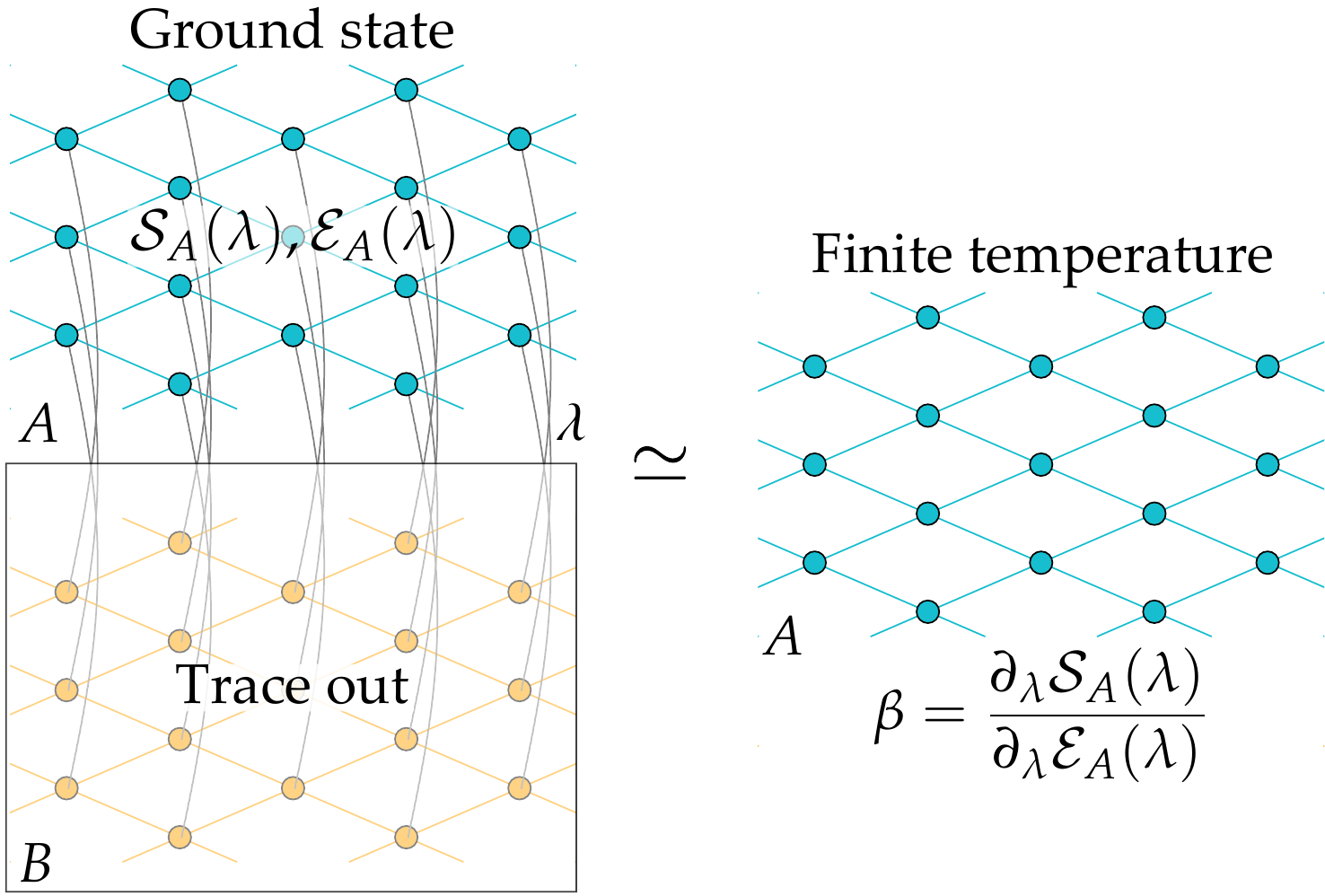}
    \caption{
      Schematic figure featuring that a
      thermal equilibrium at temperature $T=1/\beta$ (right) 
      emerges in a bipartitioned subsystem of a pure ground state by quantum entanglement 
      controlled with $\lambda$ (left). This implies that the quantum fluctuation mimics the thermal fluctuation. 
      \label{fig.schematic}}    
  \end{center}
\end{figure}

Excellent collapse of different quantities, 
$\mathcal{S}_{A}(\lambda) \simeq S_{A}(\mathcal{B}_A(\lambda))$ and 
$\mathcal{E}_{A}(\lambda) \simeq E_{A}(\mathcal{B}_A(\lambda))$,
implies that the relation  
\begin{equation}
  \hat{\rho}_{A}^{\rm red}(\lambda)
  \simeq
  \hat{\rho}_{A}^{\rm can}(\mathcal{B}_A(\lambda))   
  \label{eq.rhorho}
\end{equation}
holds between the reduced density matrix operator $\hat{\rho}_{A}^{\rm red}(\lambda)$ 
and the thermodynamic density matrix operator $\hat{\rho}_{A}^{\rm can}(\beta) $, 
independently of the detail of the subsystem $B$
whose degrees of freedom are traced out, 
as schematically shown in Fig.~\ref{fig.schematic}. 
To quantify the similarity between
these two density matrix operators, 
we calculate the fidelity $F$ of density matrix
operators $\hat{\rho}$ and $\hat{\sigma}$
on $\mathcal{H}_{A}$ defined as  
\begin{equation}
  F(\hat{\rho},\hat{\sigma})=\left({\rm Tr}_{A}\sqrt{\sqrt{\hat{\rho}}\hat{\sigma} \sqrt{\hat{\rho}}}\right)^2 
\end{equation}
for
$\hat{\rho}=\hat{\rho}_A^{\rm can}(\mathcal{B}_A (\lambda))$ and
$\hat{\sigma}=\hat{\rho}_A^{\rm red}(\lambda)$.
Note that the fidelity satisfies 
$F(\hat{\rho},\hat{\sigma})=1$ if and only if 
$\hat{\rho}=\hat{\sigma}$, and generally
$0\leqslant F(\hat{\rho},\hat{\sigma}) \leqslant 1$
~\cite{Gilchrist2005,Zhang2020}. 

\begin{figure}
  \begin{center}
    \includegraphics[width=0.95\columnwidth]{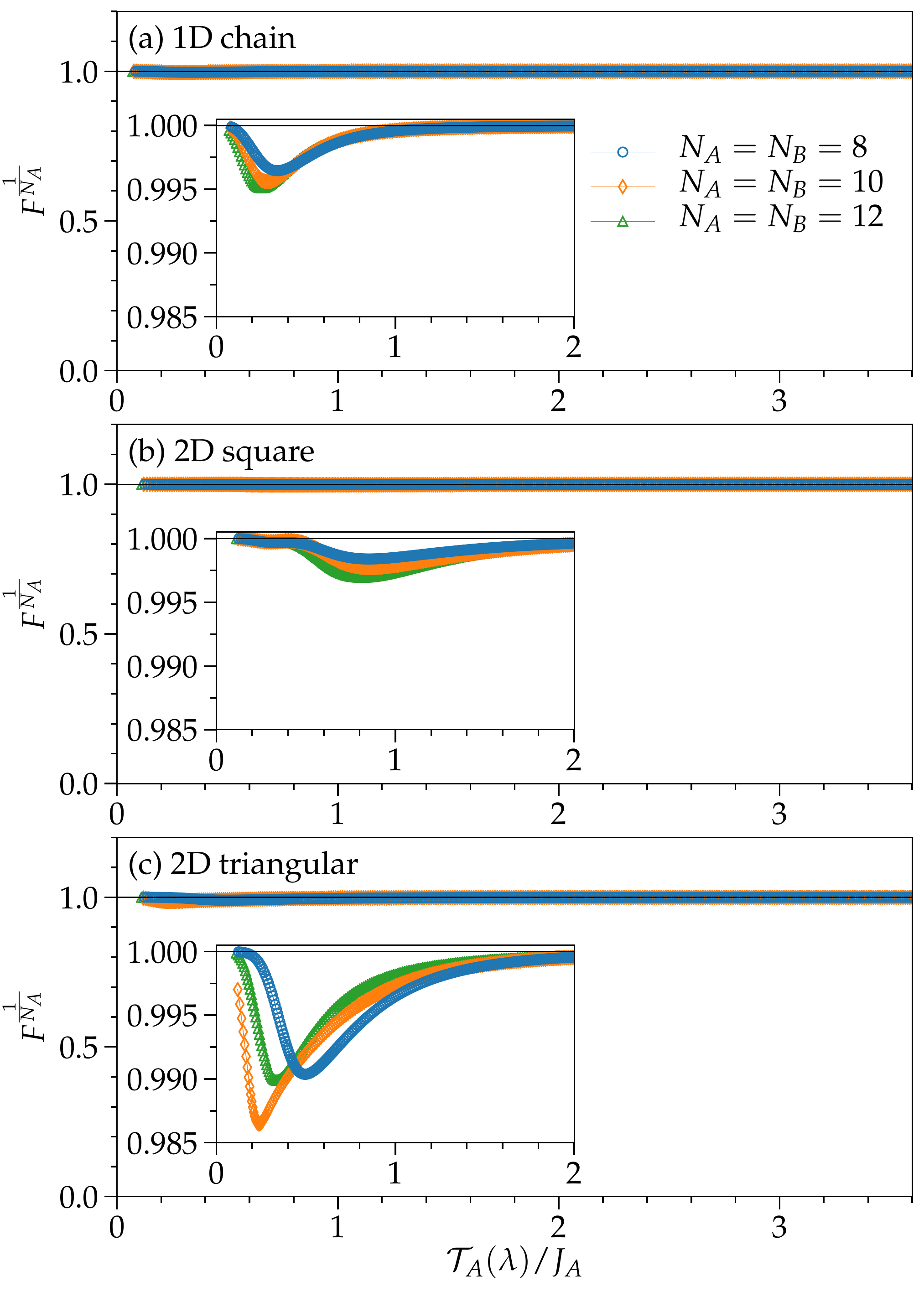}
    \caption{
      The fidelity per site
      $F(\hat{\rho},\hat{\sigma})^{\frac{1}{N_A}}$ for 
      $\hat{\rho}=\hat{\rho}_A^{\rm can}(\mathcal{B}_A (\lambda))$ and
      $\hat{\sigma}=\hat{\rho}_A^{\rm red}(\lambda)$ 
      as a function of the effective temperature $\mathcal{T}_{A}(\lambda)$ 
      calculated for (a) the two coupled 1D chains (i.e., two-leg ladder), 
      (b) the two coupled 2D square lattices (i.e., bilayer square lattice), and 
      (c) the two coupled 2D triangular lattices (i.e., bilayer triangular lattice) 
      with $N_A=N_B=8, 10,$ and $12$. 
      The insets show enlarged
      plots for $\mathcal{T}_{A}(\lambda)/J_A\leqslant2$
      and $F^{\frac{1}{N_A}}\simeq 1$.
      $J_B/J_A=1$ is assumed.
      \label{fig.fidelity}}
  \end{center}
\end{figure}

Figure~\ref{fig.fidelity} shows the fidelity
per site, $F^{\frac{1}{N_A}}$,
calculated for the three different lattice structures with $J_B/J_A=1$.  
As expected from the discussion in Sec.~\ref{sec.zeroT} and Sec.~\ref{sec.infiniteT}, 
the fidelity tends to $1$ in the limits of
$\mathcal{T}_{A}(\lambda)\to 0$ and
$\mathcal{T}_{A}(\lambda)\to \infty$.
More interestingly, 
the fidelity per site is kept large
(at least larger than $0.985$) even at
intermediate $\mathcal{T}_{A}(\lambda)$, 
verifying Eq.~(\ref{eq.rhorho}) quantitatively
(see insets of Fig.~\ref{fig.fidelity}).  
However, except for the triangular lattice,
the fidelity per site
tends to become smaller with increasing the system size. 
Obviously, calculations with larger clusters are
desirable to further examine the finite-size effects,
but currently are not feasible due to the exponentially large computational cost. 
We note that the 
fidelity with $J_B/J_A=0.5$ and $J_B/J_A=1.5$
does not significantly differ from that with $J_B/J_A=1$.

While the fidelity deviates from 1 most significantly in the intermediate temperatures, the differences
in $\mathcal{S}_A(\lambda)$ and  $S_A(\mathcal{B}_A(\lambda))$ and
in $\mathcal{E}_A(\lambda)$ and  $E_A(\mathcal{B}_A(\lambda))$
are not much visible, at least in the scale shown in Fig.~\ref{SandE}. 
To examine how the deviation of the fidelity from 1 
might reflect on microscopic observables, 
here we calculate the spin correlation function
in the subsystem $A$ 
\begin{equation}
  C_{ij}(\hat{\rho})={\rm Tr}_A \left[\hat{\rho} \hat{S}_i^z \hat{S}_j^z \right]
\end{equation}
for $\hat{\rho}=\hat{\rho}_A^{\rm red} (\lambda)$ and 
$\hat{\rho}=\hat{\rho}_A^{\rm can}(\mathcal{B}_A(\lambda))$, 
where 
$\hat{S}_{i}^z$ is the $z$ component of spin $\bs{\hat{S}}_{i}$ at site $i \in A$. 
Note that the nearest-neighbor spin correlation function
is essentially equivalent to the energy
$\mathcal{E}_A(\lambda)/N_A$ or  
$E_A(\mathcal{B}_A(\lambda))/N_A$ because the Hamiltonian $\hat{H}_A$
includes only the nearest-neighbor interactions. 

\begin{figure}
  \begin{center}
    \includegraphics[width=0.9\columnwidth]{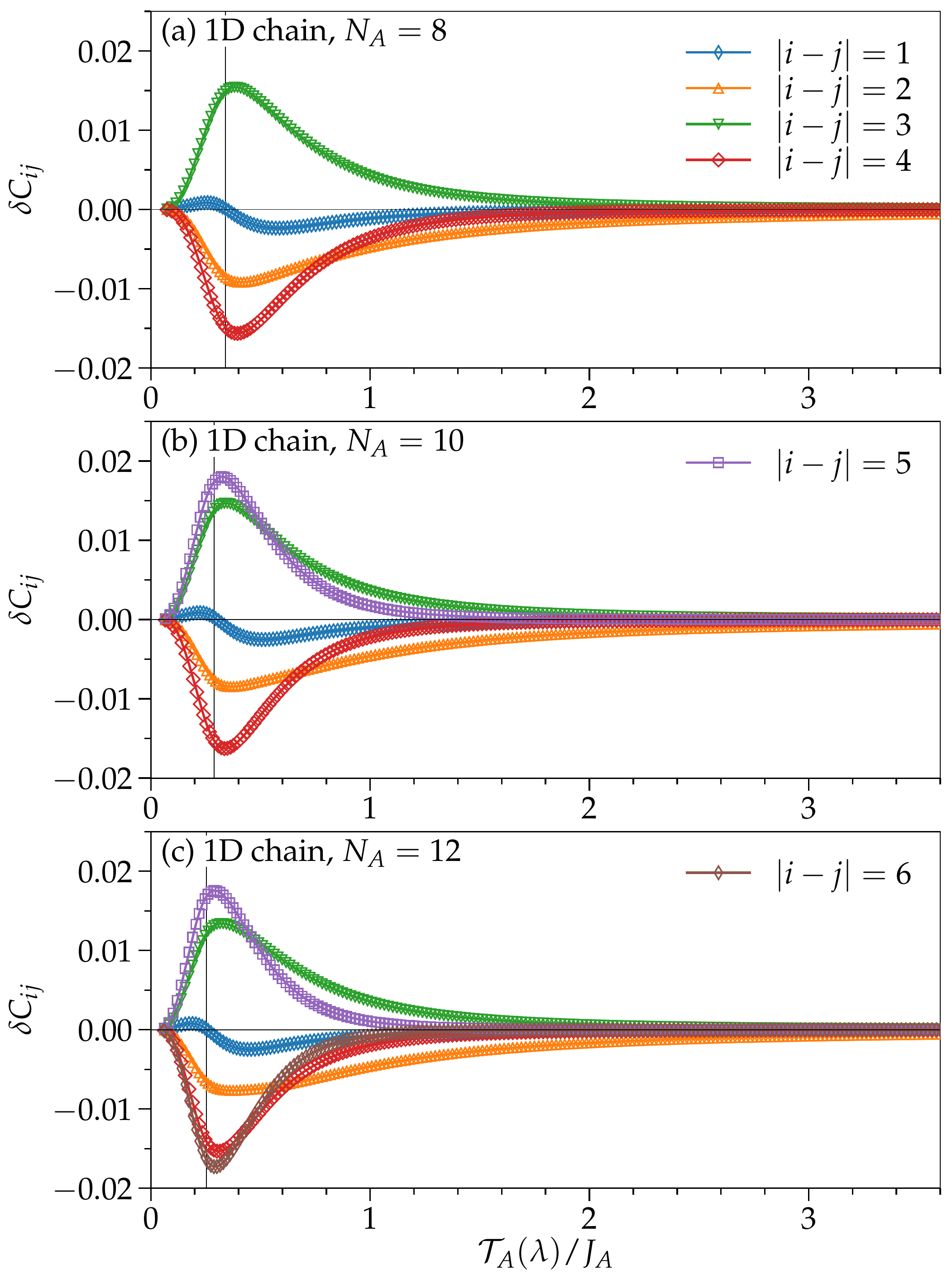}
    \caption{
      The difference of the spin correlation function $\delta C_{ij}$ 
      for various values of the spatial distance $|i-j|$ 
      as a function of the effective temperature $\mathcal{T}_{A}(\lambda)$.      
      The results are obtained for the 1D chain with 
      (a) $N_A=8$,
      (b) $N_A=10$, and
      (c) $N_A=12$.
      The vertical line indicates the temperature
      where the fidelity $F$ deviates most from 1 in Fig.~\ref{fig.fidelity}(a).
      \label{fig.correlation}}    
  \end{center}
\end{figure}

\begin{figure}
  \begin{center}
    \includegraphics[width=0.9\columnwidth]{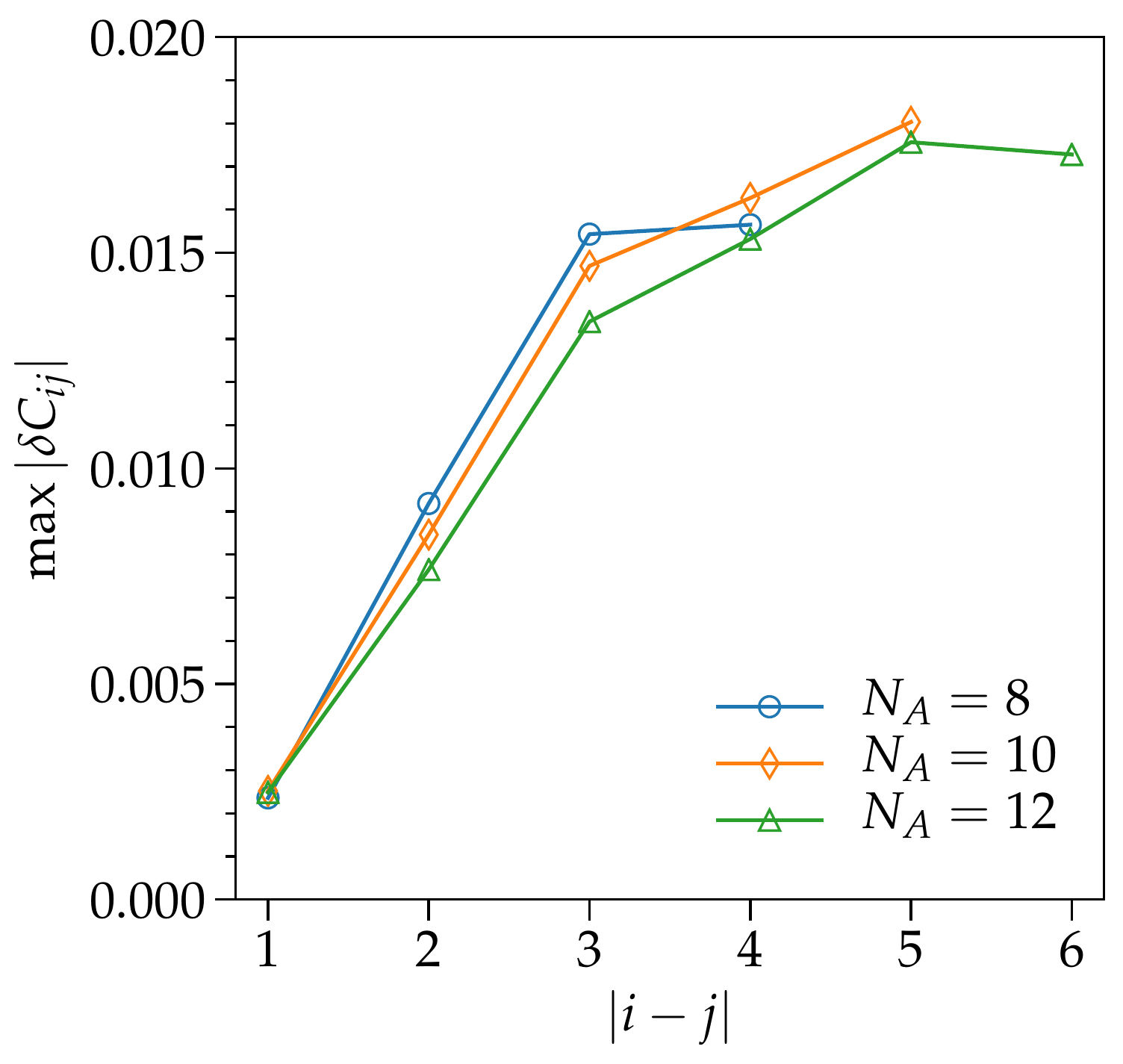}
    \caption{
      The maximum of the absolute difference of the spin correlation functions,
      $\max |\delta C_{ij}|$, among different temperatures shown in Fig.~\ref{fig.correlation} as a function of
      the spatial distance $|i-j|$ for the 1D chain with $N_A=$8, 10, and 12.
      Due to the translational symmetry, only the results for
      $1 \leqslant |i-j| \leqslant N_A/2$ are shown. 
      \label{fig.maxdC}}    
  \end{center}
\end{figure}

Figure~\ref{fig.correlation} shows the difference of the spin correlation function
\begin{equation}
  \delta C_{ij}(\lambda) = C_{ij}(\hat{\rho}_{A}^{\rm can}(\mathcal{B}_A(\lambda))) -  C_{ij}(\hat{\rho}_{A}^{\rm red}(\lambda))  
\end{equation}
as a function of $\mathcal{T}_A(\lambda)$ with $J_B/J_A=1$.
Here we focus on the 1D systems with $N_A=8, 10,$ and $12$ 
to find a systematic dependence on the system size.  
Due to the periodic-boundary conditions,
$\delta C_{ij}$ for the 1D systems depends only
on the spatial distance $|i-j|$ between sites $i$ and $j$. 
It is found that 
the difference of the next-nearest neighbor and longer-range ($|i-j|\geqslant 2$)
spin correlation functions 
is pronounced in the temperature range where the fidelity exhibits a dip. 
On the other hand, the difference in the nearest-neighbor ($|i-j|=1$)
correlation function is less significant, as expected from the energy calculations. 
It is also found that, 
the maximum of the absolute difference of the spin correlation function,
$\max |\delta C_{ij}(\lambda)|$, among different temperatures 
remains almost unchanged for $|i-j|=1$ and
even tends to decrease for $|i-j|\geqslant2$ 
with increasing the size $N_A$, 
in spite of the decrease of the fidelity with increasing $N_A$
(see Fig.~\ref{fig.maxdC}). 
These results suggest that 
the difference between the two density matrix operators 
$\hat{\rho}_A^{\rm red} (\lambda)$ and 
$\hat{\rho}_A^{\rm can}(\mathcal{B}_A(\lambda))$ in the intermediate temperatures 
can cause a rather prominent 
signature in microscopic quantities such as the
longer-range spin correlation functions that do not
contribute to the energy.

Although the strong finite-size effect do not allow us
to perform a systematic analysis for the 2D square and triangular lattices,
we have also found that, in the temperature range where
the fidelity deviates from 1, the spin correlation functions
beyond the nearest-neighbor distance tend to differ most in the 2D systems.

\section{Two analytical examples}~\label{sec:analytical}
To support the numerical finding in Sec.~\ref{sec.results},
here we consider two examples, 
free bosons and free fermions under pairing field,  
which can be solved analytically, and show that
the reduced density matrix operator of a partitioned 
subsystem of a ground state is identical to
the thermodynamic density matrix operator, provided 
that the effective temperature is properly introduced as in Eq.~(\ref{beta_dSdE}).

\subsection{Bosons under pairing field}
First we analyze the entanglement between
free bosons under pairing field by considering 
the Hamiltonian of the form in Eq.~(\ref{fullHam})
with 
\begin{alignat}{1}
  &\hat{H}_A  = \omega_A \left(\hat{a}^\dag \hat{a} + \frac{1}{2}\right),\\
  &\hat{H}_B  = \omega_B \left(\hat{b}^\dag \hat{b} + \frac{1}{2}\right),
\end{alignat}
and 
\begin{alignat}{1}
  &\hat{V}_{AB}(\lambda)  = \lambda \left(\hat{a}\hat{b}+\hat{b}^\dag \hat{a}^\dag\right),   
\end{alignat}
where $\hat{a}$ and $\hat{b}$ are boson annihilation operators 
on $\mathcal{H}_A$ and $\mathcal{H}_B$, respectively.  
The operators satisfy the commutation relations 
$[\hat{a},\hat{a}^\dag]=1$,
$[\hat{b},\hat{b}^\dag]=1$,
$[\hat{a},\hat{b}^\dag]=0$, and
$[\hat{a},\hat{b}]=0$.
We assume that $\omega_A >0$, $\omega_B >0$, and
$|\lambda|<(\omega_A + \omega_B)/2$.
More precise restrictions on the parameters
are discussed after Eq.~(\ref{stable}).

By introducing new bosonic operators
$\hat{\alpha}$ and $\hat{\beta}$ via a Bogoliubov transformation as 
\begin{equation}
  \begin{bmatrix}
    \hat{\alpha}\\
    \hat{\beta}^\dag
  \end{bmatrix}
  =
  \begin{bmatrix}
    \cosh\theta & \sinh\theta\\
    \sinh\theta & \cosh\theta
  \end{bmatrix}
  \begin{bmatrix}
    \hat{a}\\
    \hat{b}^\dag
  \end{bmatrix}
\end{equation}
with $\theta$ satisfying that 
\begin{equation}
  \lambda= \omega\tanh(2\theta) \label{lambda}
\end{equation}
and
\begin{equation}
  \omega=\frac{\omega_A+\omega_B}{2}, 
\end{equation}
the Hamiltonian $\hat{H}(\lambda)$ can be diagonalized as
\begin{equation}
  \hat{H}(\lambda)=
  \Omega_{\alpha} \hat{\alpha}^\dag \hat{\alpha}
  +\Omega_{\beta} \hat{\beta}^\dag \hat{\beta}
  +E_0, 
\end{equation}
where
\begin{alignat}{1}
  &\Omega_{\alpha} = \frac{\omega_A\cosh^2{\theta}-\omega_B \sinh^2{\theta}}
        {\cosh^2{\theta}+\sinh^2{\theta}},\\
  &\Omega_{\beta} = \frac{\omega_B\cosh^2{\theta}-\omega_A \sinh^2{\theta}}
              {\cosh^2{\theta}+\sinh^2{\theta}},
\end{alignat}
and
\begin{alignat}{1}
  &E_0
  =\frac{\omega}{\cosh^2{\theta}+\sinh^2{\theta}}
  =\frac{\Omega_\alpha+\Omega_\beta}{2}.     
\end{alignat}

Let us assume that 
\begin{equation}
  \Omega_\alpha>0 \text{\quad and \quad} \Omega_\beta>0. 
  \label{stable}
\end{equation}
These inequalities are satisfied for any $\theta$ if $\omega_A=\omega_B$. 
However, if $\omega_A\not=\omega_B$, 
these inequalities are satisfied only in a limited range of $\theta$.
For example, if $\omega_B/\omega_A < 1$, 
$\Omega_\alpha>0$ is satisfied for any $\theta$ but 
$\Omega_\beta>0$ is satisfied only if 
$\tanh^2{\theta}<\omega_B/\omega_A<1$.
A similar condition can be found for $\omega_A/\omega_B < 1$. 
Below we only consider the parameter region that satisfies 
the inequalities in Eq.~(\ref{stable}).

Since
$\Omega_\alpha>0$ and $\Omega_\beta>0$,  
the ground state $|\Psi_0(\lambda)\rangle$
of $\hat{H}(\lambda)$ should be a vacuum state 
of bosons $\hat{\alpha}$ and $\hat{\beta}$ 
satisfying 
$\hat{\alpha}|\Psi_0 (\lambda)\rangle=0$ and
$\hat{\beta}|\Psi_0 (\lambda)\rangle=0$.
Using the vacuum states $|0\rangle_A$ and $|0\rangle_B$
in $\mathcal{H}_A$ and $\mathcal{H}_B$, respectively satisfying 
$\hat{a}|0\rangle_A=0$ and $\hat{b}|0\rangle_B=0$, 
the ground state can be given explicitly as
\begin{alignat}{1}
  |\Psi_0 (\lambda)\rangle
  &= \frac{1}{\cosh{\theta}} \e^{-(\tanh{\theta})\hat{a}^\dag\hat{b}^\dag}
  |0\rangle_A |0\rangle_B \notag \\
  &=\frac{1}{\cosh{\theta}}\sum_{n=0}^{\infty}(-\tanh{\theta})^n |n\rangle_A |n\rangle_B,
  \label{GSHO}
\end{alignat}
with
$ |n\rangle_A=(n!)^{-1/2}(\hat{a}^\dag)^n |0\rangle_A$ and 
$ |n\rangle_B=(n!)^{-1/2}(\hat{b}^\dag)^n |0\rangle_B$.
The entangled state of the form in 
Eq.~(\ref{GSHO}) has several applications 
including the Unruh effect~\cite{Unruh1976,Crispino2008} and
a two-mode squeezed state in quantum optics~\cite{Lee1990}.

\begin{figure*}
  \begin{center}
    \includegraphics[width=1.95\columnwidth]{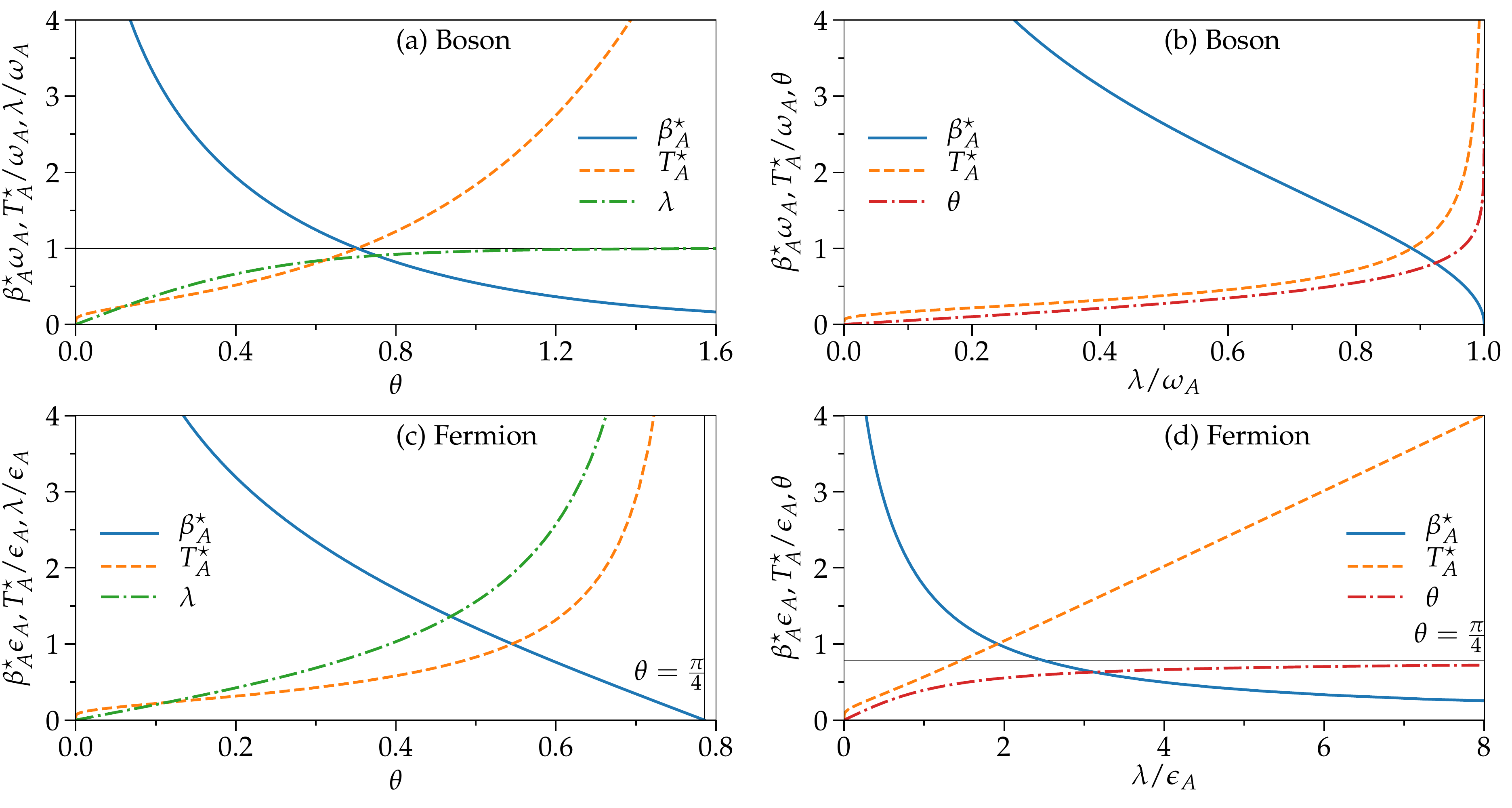}
    \caption{
      $\beta_A^\star$, $T_A^\star\,(=1/\beta_A^\star)$, and $\lambda$ as a function of
      $\theta$ for (a) the bosonic system with $\omega_A=\omega_B$ and 
      (c) the fermionic system with $\epsilon_A=\epsilon_B$. 
      $\beta_A^\star$, $T_A^\star\,(=1/\beta_A^\star)$, and $\theta$ as a function of
      $\lambda$ for (b) the bosonic system with $\omega_A=\omega_B$ and 
      (d) the fermionic system with $\epsilon_A=\epsilon_B$. 
      The horizontal line in (a) indicates the asymptote of
      $\lim_{\theta\to\infty}\lambda(\theta)/\omega_A=1$, 
      the vertical line in (c) indicates $\theta=\pi/4$, and
      the horizontal line in (d) indicates the asymptote of 
      $\lim_{\lambda\to\infty}\theta(\lambda)=\pi/4$.  
      \label{beta_T_BF}}     
  \end{center}
\end{figure*}

By tracing out the degrees of freedom in $\mathcal{H}_{B}$
from the ground-state density matrix operator $|\Psi_0\rangle \langle \Psi_0|$, 
we obtain the reduced density matrix operator of subsystem $A$:  
\begin{equation}
  \hat{\rho}_A^{\rm red}(\lambda)=\frac{1}{\cosh^2{\theta}}
  \sum_{n=0}^{\infty} (\tanh^2{\theta})^n |n\rangle_{A} {_{A} \langle n|}.
  \label{rhoredHO}
\end{equation}
On the other hand, by noticing
$\hat{H}_{A}|n\rangle_{A}=\omega_A (n+\frac{1}{2})|n\rangle_{A}$, 
the thermodynamic density matrix operator of the isolated subsystem $A$ is given by 
\begin{eqnarray}
  \hat{\rho}_A^{\rm can}(\beta) & =&
  (1 - \e^{-\beta \omega_A})
  \sum_{n=0}^{\infty} \e^{-\beta \omega_A n}  |n\rangle_{A} {_{A} \langle n|}  
  \label{rhothermHO} \\
  & = &  (\e^{ \beta \omega_A/2} - \e^{-\beta \omega_A/2})
  \sum_{n=0}^{\infty} \e^{-\beta \omega_A  \left( n + \frac{1}{2} \right)  }  |n\rangle_{A} {_{A} \langle n|}.  
  \label{eq:rhothermHO}
\end{eqnarray}
By comparing Eq.~(\ref{rhoredHO}) with Eq.~(\ref{rhothermHO}), 
it is found that $\hat{\rho}_A^{\rm red}(\lambda)$ 
is exactly the same as $\hat{\rho}_A^{\rm can}(\beta)$ when $\beta=\beta_A^\star$ with
\begin{alignat}{1}
  \beta_A^\star
  &=-\frac{1}{\omega_A}\ln{\tanh^2{\theta}}\label{betaHO} \\ 
  &=-\frac{1}{\omega_A}\ln
  \left[\frac{\omega}{\lambda}\pm\sqrt{\left(\frac{\omega}{\lambda}\right)^2-1} \right]^2, 
\end{alignat}
where $+$ ($-$) sign is taken for $\theta>0$ ($\theta<0$).
Figures~\ref{beta_T_BF}(a) and \ref{beta_T_BF}(b) show $\theta$ and $\lambda$ dependence of
$\beta_A^\star$ and $T_A^\star=1/\beta_A^\star$, respectively. 
We can furthermore find that the entanglement Hamiltonian
$\hat{\mathcal{I}}^{\rm red}_A= -\ln \hat{\rho}_A^{\rm red}$
is proportional to $\hat{H}_A$ with coefficient $\beta_A^{\star}$:
\begin{equation}
  \hat{\mathcal{I}}^{\rm red}_A  
  = \beta_A^{\star} \hat{H}_{A} + \frac{1}{2}\ln Z_A^2,  
\end{equation}
where $Z_A^2=\cosh^2{\theta}\sinh^2{\theta} = \left(\e^{\beta_A^\star \omega_A/2}-\e^{-\beta_A^\star \omega_A/2}\right)^{-2}$ 
and the spectral representation of the number operator 
$\hat{a}^\dag \hat{a}=\sum_{n=0}^\infty n|n\rangle_A {_A}\langle n|$ is used.
Next, we shall show that
the inverse temperature $\beta^\star_A$ given in Eq.~(\ref{betaHO}) is the same as the effective inverse temperature 
$\mathcal{B}_A(\lambda)=
\partial_{\lambda}{\mathcal{S}_A}/
\partial_{\lambda}{\mathcal{E}_A}$ introduced in Eq.~(\ref{beta_dSdE}).

The effective
inverse temperature $\mathcal{B}_{A}(\lambda)$ is evaluated 
from the entanglement von Neumann entropy $\mathcal{S}_{A}(\lambda)$
and the energy $\mathcal{E}_{A}(\lambda)$ of subsystem $A$ for the ground state $|\Psi_0 (\lambda)\rangle$. 
Equation~(\ref{rhoredHO}) implies that 
$\hat{\rho}_A^{\rm red}(\lambda)$ contains 
the eigenstate $|n\rangle_A$ of $\hat{H}_A$
with the probability
\begin{equation}
  p_n=\frac{(\tanh^2\theta)^n}{\cosh^2\theta}. 
\end{equation}
Therefore, the entanglement von Neumann entropy $\mathcal{S}_{A}(\lambda)$ is calculated as 
\begin{alignat}{1}
  \mathcal{S}_{A}(\lambda)
  &= -{\rm Tr}_A \left[\hat{\rho}_A^{\rm red}(\lambda) \ln \hat{\rho}_A^{\rm red}(\lambda) \right] \notag \\
  &= - \sum_{n=0}^{\infty} p_n \ln p_n \notag \\
  &=(\cosh^2\theta) \ln {\cosh^2{\theta}}
  -(\sinh^2\theta) \ln {\sinh^2{\theta}},
  \label{SHO}
\end{alignat}
where $\sum_{n=0}^{\infty}n x^n=x/(1-x)^2$ for $|x|<1$ is used 
in the last equality. 
Similarly, the energy $\mathcal{E}_{A}(\lambda)$ is calculated as
\begin{alignat}{1}
  \mathcal{E}_{A}(\lambda)
  &={\rm Tr}_A\left[\hat{\rho}_A^{\rm red}(\lambda)\hat{H}_A\right] \notag \\
  &=\omega_A\sum_{n=0}^{\infty} p_n \left(n+\frac{1}{2}\right)\notag \\
  &=\omega_A \left(\sinh^2{\theta} + \frac{1}{2} \right),
  \label{EHO}
\end{alignat}
where $\sum_{n=0}^{\infty}p_n=1$ is used. 
We thus find that
$\partial_\theta \mathcal{S}_A=-(2\cosh\theta \sinh\theta) \ln \tanh^2\theta$ and
$\partial_\theta \mathcal{E}_A= 2\omega_A\cosh\theta \sinh\theta$.
Therefore, the effective inverse temperature $\mathcal{B}_A(\lambda)$ is calculated as 
\begin{equation}
  \mathcal{B}_A(\lambda)=
  \frac{\partial_\lambda \theta \ \partial_\theta \mathcal{S}_A}
       {\partial_\lambda \theta \ \partial_\theta \mathcal{E}_A}
       =-\frac{1}{\omega_A} \ln \tanh^2{\theta}
       =\beta_A^\star.
       \label{BHO}
\end{equation}
From Eqs.~(\ref{rhoredHO})--(\ref{betaHO}) and (\ref{BHO}), we can conclude that 
the relation 
\begin{equation}
  \hat{\rho}_{A}^{\rm can}(\mathcal{B}_A(\lambda)) = \hat{\rho}_{A}^{\rm red}(\lambda)
  \label{can_red}
\end{equation}
holds exactly and thus is no longer a conjecture in this case.

We note that the relation in Eq.~(\ref{BHO}) can also be
derived by directly
calculating $\partial \mathcal{S}_{A}/\partial \mathcal{E}_{A}$. 
Namely, it follows from Eq.~(\ref{EHO}) that 
$\sinh^2{\theta}=\frac{\mathcal{E}_A}{\omega_A}-\frac{1}{2}$ and 
$\cosh^2{\theta}=\frac{\mathcal{E}_A}{\omega_A}+\frac{1}{2}$. 
Therefore, $\mathcal{S}_{A}$ can be expressed in terms of $\mathcal{E}_{A}$ as
\begin{equation}
  \mathcal{S}_{A} =
  \left(\frac{\mathcal{E}_A}{\omega_A}+\frac{1}{2}\right) \ln
  \left(\frac{\mathcal{E}_A}{\omega_A}+\frac{1}{2}\right)
  -
  \left(\frac{\mathcal{E}_A}{\omega_A}-\frac{1}{2}\right) \ln
  \left(\frac{\mathcal{E}_A}{\omega_A}-\frac{1}{2}\right),  
\end{equation}
and thus one can readily show that  
\begin{equation}
  \frac{\partial \mathcal{S}_A}{\partial \mathcal{E}_A}
  =\beta_A^\star.  
\end{equation}

To confirm more specifically the correspondence
between the two density matrix operators in
Eq.~(\ref{can_red}),  
let us rewrite
the entanglement von Neumann entropy $\mathcal{S}_A$ and
the energy $\mathcal{E}_A$
in terms of $\beta_A^\star$, instead of $\theta$. 
It follows from Eq.~(\ref{betaHO}) that
\begin{alignat}{1}
  &\tanh^2{\theta}=\e^{-\beta_A^\star \omega_A},\\
  &\cosh^2\theta=1+n_{\beta_A^\star}(\omega_A), \label{cosh2}
\end{alignat}
and
\begin{alignat}{1}
  &\sinh^2\theta=n_{\beta_A^\star}(\omega_A),     \label{sinh2}
\end{alignat}
where
\begin{equation}
  n_{\beta_A^\star}(\omega_A)= \frac{1}{\e^{\beta_A^\star \omega_A}-1} 
\end{equation}
is the Bose-Einstein distribution function at inverse temperature $\beta_A^\star$.
Substituting Eqs.~(\ref{cosh2}) and (\ref{sinh2})
into Eqs.~(\ref{SHO}) and (\ref{EHO}) yields that 
\begin{alignat}{1}
  \mathcal{S}_{A} &= (1+n_{\beta_A^\star})
  \ln(1+n_{\beta_A^\star})
  -n_{\beta_A^\star} \ln n_{\beta_A^\star} 
\end{alignat}
and
\begin{alignat}{1}
  \mathcal{E}_{A} &= \omega_A \left(n_{\beta_A^\star}+\frac{1}{2}\right), 
\end{alignat}
which are familiar forms of the thermodynamic 
entropy and the internal energy of
free bosons, respectively.  
One can also readily find that
the positive square root $Z_A=\left(\e^{\beta^\star_A\omega_A/2}-\e^{-\beta^\star_A\omega_A/2}\right)^{-1}>0$
of $Z_A^2$ gives the corresponding partition function, i.e.,
$Z_A={\rm Tr}_{A}\left[ \e^{-\beta^\star_A \hat{H}_A} \right]$. 

By doing the same analysis for subsystem $B$, 
one can find the relation between the effective temperatures
$\mathcal{T}_A(\lambda)=\mathcal{B}_A^{-1}(\lambda)$ and 
$\mathcal{T}_B(\lambda)=\mathcal{B}_B^{-1}(\lambda)$ as 
\begin{equation}
  \mathcal{T}_{A}(\lambda)/\omega_A=\mathcal{T}_{B}(\lambda)/\omega_B.
  \label{eq:temp_B}
\end{equation} 
Finally, we note that all these analyses given above are based on the ground state
in Eq.~(\ref{GSHO}) under the conditions in Eq.~(\ref{stable}). 
One can readily show that if $\omega_A\not = \omega_B$,
the maximum of the effective temperature is bounded.
For example, when 
$\omega_B/\omega_A<1$, 
$\tanh^2{\theta}<\omega_B/\omega_A$ should be satisfied 
in order to satisfy the conditions in Eq.~(\ref{stable}).
This implies that 
$ \ln{(\omega_A/\omega_B)} < \beta_A^\star \omega_A<\infty$,
or equivalently 
$ 0 < T_A^\star/\omega_A< 1/\ln{(\omega_A/\omega_B)}$.

\subsection{Fermions under pairing field}
Next we analyze the entanglement between
free fermions under pairing field by considering
the Hamiltonian of the form in Eq.~(\ref{fullHam})
with 
\begin{alignat}{1}
  &\hat{H}_A  = \epsilon_A \left(\hat{a}^\dag \hat{a}-\frac{1}{2}\right),\label{FHA}\\
  &\hat{H}_B  = \epsilon_B \left(\hat{b}^\dag \hat{b}-\frac{1}{2}\right),\label{FHB}
\end{alignat}
and
\begin{alignat}{1}
  &\hat{V}_{AB}(\lambda)  = \lambda \left(\hat{a}\hat{b}+\hat{b}^\dag \hat{a}^\dag\right),   \label{FVAB}
\end{alignat}
where $\hat{a}$ and $\hat{b}$ are fermion annihilation operators 
on $\mathcal{H}_A$ and $\mathcal{H}_B$, respectively.
The subtraction of $1/2$ in Eqs.~(\ref{FHA}) and (\ref{FHB})
is made simply to find a formal similarity with
the bosonic case. 
The operators satisfy the anticommutation relations 
$\{\hat{a},\hat{a}^\dag\}=1$,
$\{\hat{b},\hat{b}^\dag\}=1$,
$\{\hat{a},\hat{b}^\dag\}=0$, and
$\{\hat{a},\hat{b}\}=0$.
Here 
we assume that $\epsilon_A >0$ and $\epsilon_B >0$.
An interpretation of this assumption is, for example, 
that we consider 
the coupling between two fermion particles (holes) added 
above (below) the Fermi sea,
with the energies $\epsilon_A$ and
$\epsilon_B$ measured from the Fermi level. 
More precise restrictions on the parameters
are discussed after Eq.~(\ref{Fstable}).

By introducing new fermionic operators
$\hat{\alpha}$ and $\hat{\beta}$ via a Bogoliubov transformation as 
\begin{equation}
  \begin{bmatrix}
    \hat{\alpha}\\
    \hat{\beta}^\dag
  \end{bmatrix}
  =
  \begin{bmatrix}
    \cos\theta & -\sin\theta\\
    \sin\theta & \cos\theta
  \end{bmatrix}
  \begin{bmatrix}
    \hat{a}\\
    \hat{b}^\dag
  \end{bmatrix}
\end{equation}
with $\theta$ satisfying that 
\begin{alignat}{1}
  \lambda&= \epsilon\tan(2\theta)\label{Flambda}
\end{alignat}
and
\begin{alignat}{1}
  \epsilon&=\frac{\epsilon_A+\epsilon_B}{2}, 
\end{alignat}
the Hamiltonian $\hat{H}(\lambda)$ can be diagonalized as
\begin{equation}
  \hat{H}(\lambda)=
  \xi_{\alpha} \hat{\alpha}^\dag \hat{\alpha}
  +\xi_{\beta} \hat{\beta}^\dag \hat{\beta}
  +E_0, 
\end{equation}
where
\begin{alignat}{1}
  &\xi_{\alpha} = \epsilon_A\cos^2{\theta}-\epsilon_B \sin^2{\theta},\\
  &\xi_{\beta}  = \epsilon_B\cos^2{\theta}-\epsilon_A \sin^2{\theta},
\end{alignat}
and
\begin{alignat}{1}
  &E_0
  =-\epsilon(\cos^2{\theta}-\sin^2{\theta})
  =-\frac{\xi_\alpha+\xi_\beta}{2}.
\end{alignat}

Similarly to the bosonic case, we assume that
\begin{equation}
  \xi_\alpha>0 \text{\quad and \quad} \xi_\beta>0. 
  \label{Fstable}
\end{equation}
If $\epsilon_A=\epsilon_B$,
these inequalities are satisfied for
$-\frac{\pi}{4}<\theta < \frac{\pi}{4}$, 
implying that $-\infty < \lambda < \infty$. 
However, if $\epsilon_A\not=\epsilon_B$, 
the range of $\theta$ and hence $\lambda$
allowed is more restricted. 
For example, if $\epsilon_B/\epsilon_A < 1$, 
$\xi_\alpha>0$ is satisfied for any $\theta$ but 
$\xi_\beta>0$ is satisfied only if 
$\tan^2{\theta}<\epsilon_B/\epsilon_A<1$.
A similar condition can be found for $\epsilon_A/\epsilon_B < 1$. 
Below we only consider the parameter region that satisfies 
the inequalities in Eq.~(\ref{Fstable}).

Since $\xi_\alpha>0$ and $\xi_\beta>0$,  
the ground state $|\Psi_0(\lambda)\rangle$
of $\hat{H}(\lambda)$ should be a vacuum state 
of fermions $\hat{\alpha}$ and $\hat{\beta}$ 
satisfying 
$\hat{\alpha}|\Psi_0 (\lambda)\rangle=0$ and
$\hat{\beta}|\Psi_0 (\lambda)\rangle=0$.
Using the vacuum states $|0\rangle_A$ and $|0\rangle_B$
in $\mathcal{H}_A$ and $\mathcal{H}_B$, respectively satisfying 
$\hat{a}|0\rangle_A=0$ and $\hat{b}|0\rangle_B=0$, 
the ground state can be given explicitly as
\begin{alignat}{1}
  |\Psi_0 (\lambda)\rangle
  &= \cos{\theta} \e^{(\tan\theta) \hat{a}^\dag\hat{b}^\dag}
  |0\rangle_A |0\rangle_B \notag \\
  &=\cos{\theta}\sum_{n=0}^{1}(\tan{\theta})^n |n\rangle_A |n\rangle_B,
  \label{GSCooper}
\end{alignat}
with
$ |n\rangle_A=(\hat{a}^\dag)^n |0\rangle_A$ and 
$ |n\rangle_B=(\hat{b}^\dag)^n |0\rangle_B$.
The multi-mode-generalization of
the entangled state of the form in 
Eq.~(\ref{GSCooper}) is the Bardeen-Cooper-Schrieffer
(BCS) wave function~\cite{BCS1957}.

By tracing out the degrees of freedom in $\mathcal{H}_{B}$
from the ground-state density matrix operator $|\Psi_0\rangle \langle \Psi_0|$, 
we obtain the reduced density matrix operator of subsystem $A$: 
\begin{equation}
  \hat{\rho}_A^{\rm red}(\lambda)=\cos^2{\theta}
  \sum_{n=0}^{1} (\tan^2{\theta})^n |n\rangle_{A} {_{A} \langle n|}.
  \label{rhoredCooper}
\end{equation}
On the other hand, by noticing
$\hat{H}_{A}|n\rangle_{A}=\epsilon_A (n-\frac{1}{2})|n\rangle_{A}$, 
the thermodynamic density matrix operator of subsystem $A$ is given by 
\begin{eqnarray}
  \hat{\rho}_A^{\rm can}(\beta) &=&
  \frac{1}{1+\e^{-\beta\epsilon_A }}
    \sum_{n=0}^{1} \e^{-\beta \epsilon_A n}  |n\rangle_{A} {_{A} \langle n|} 
  \label{rhothermCooper}\\
  &=& \frac{1}{
  \e^{ \beta\epsilon_A/2}+
  \e^{-\beta\epsilon_A/2}}
  \sum_{n=0}^{1} \e^{-\beta \epsilon_A \left(n - \frac{1}{2} \right) }  |n\rangle_{A} {_{A} \langle n|}. 
  \label{eq:rhothermCooper}
\end{eqnarray}
By comparing Eq.~(\ref{rhoredCooper}) with Eq.~(\ref{rhothermCooper}), 
it is found that $\hat{\rho}_A^{\rm red}(\lambda)$ 
is exactly the same as $\hat{\rho}_A^{\rm can}(\beta)$ when $\beta=\beta_A^\star$ with
\begin{alignat}{1}
  \beta_A^\star
  &=-\frac{1}{\epsilon_A}\ln{\tan^2{\theta}} \label{betaCooper}\\
  &=-\frac{1}{\epsilon_A}
  \ln
  \left[
    -\frac{\epsilon}{\lambda}
    \pm
    \sqrt{
      \left(\frac{\epsilon}{\lambda}\right)^2+1
    }\right]^2,
\end{alignat}
where $+$ ($-$) sign is taken for $\theta>0$ ($\theta<0$).  
Figures~\ref{beta_T_BF}(c) and \ref{beta_T_BF}(d) show $\theta$ and $\lambda$ dependence of
$\beta_A^\star$ and $T_A^\star=1/\beta_A^\star$, respectively. 
We can furthermore find that the entanglement Hamiltonian
$\hat{\mathcal{I}}^{\rm red}_A= -\ln \hat{\rho}_A^{\rm red}$
is proportional to $\hat{H}_A$ with coefficient $\beta_A^{\star}$:
\begin{equation}
  \hat{\mathcal{I}}_A^{\rm red}  
  = \beta_A^{\star} \hat{H}_{A} + \frac{1}{2}\ln Z_A^2,  
\end{equation}
where $Z_A^2=1/\cos^2{\theta}\sin^2{\theta}=\left( \e^{ \beta^\star_A\epsilon_A/2}+\e^{-\beta^\star_A\epsilon_A/2} \right)^2$ 
and the spectral representation of the number operator
$\hat{a}^\dag \hat{a}=\sum_{n=0}^1 n|n\rangle_A {_A}\langle n|$ is used.
Next, we shall show that
the inverse temperature $\beta^\star_A$ given in Eq.~(\ref{betaCooper}) is the 
same as the effective inverse temperature 
$\mathcal{B}_A(\lambda)=
\partial_{\lambda}{\mathcal{S}_A}/
\partial_{\lambda}{\mathcal{E}_A}$ introduced in Eq.~(\ref{beta_dSdE}).

The effective inverse temperature $\mathcal{B}_{A}$ is evaluated from the
entanglement von Neumann entropy $\mathcal{S}_{A}(\lambda)$
and the energy $\mathcal{E}_{A}(\lambda)$ of subsystem $A$ for the ground state $|\Psi_0 (\lambda)\rangle$. 
Equation~(\ref{rhoredCooper}) implies that 
$\hat{\rho}_A^{\rm red}(\lambda)$ contains 
the eigenstate $|n\rangle_A$ of $\hat{H}_A$
with the probability
\begin{equation}
  p_n=\cos^2\theta (\tan^2\theta)^n, 
\end{equation}
or more explicitly $p_0=\cos^2\theta$ and $p_1=\sin^2\theta$.
Therefore, the entanglement von Neumann entropy $\mathcal{S}_{A}(\lambda)$ is calculated as 
\begin{alignat}{1}
  \mathcal{S}_{A}(\lambda)
  &= -{\rm Tr}_A \left[\hat{\rho}_A^{\rm red}(\lambda) \ln \hat{\rho}_A^{\rm red}(\lambda) \right] \notag \\
  &= - \sum_{n=0}^{1} p_n \ln p_n \notag \\
  &=-(\cos^2\theta) \ln {\cos^2{\theta}}
  -(\sin^2\theta) \ln {\sin^2{\theta}},
  \label{SCooper}
\end{alignat}
Similarly, the energy $\mathcal{E}_{A}(\lambda)$ is calculated as
\begin{alignat}{1}
  \mathcal{E}_{A}(\lambda)
  &={\rm Tr}_A\left[\hat{\rho}_A^{\rm red}(\lambda)\hat{H}_A\right] \notag \\
  &=\epsilon_A\sum_{n=0}^{1} p_n \left(n-\frac{1}{2}\right)\notag \\
  &=\epsilon_A \left(\sin^2{\theta} - \frac{1}{2} \right).
  \label{ECooper}
\end{alignat}
We thus find that
$\partial_\theta \mathcal{S}_A= (2\cos\theta \sin\theta) \ln \tan^2\theta$ and
$\partial_\theta \mathcal{E}_A= -2\epsilon_A\cos\theta \sin\theta$.
Therefore, the effective inverse temperature $\mathcal{B}_A(\lambda)$ is calculated as 
\begin{equation}
  \mathcal{B}_A(\lambda)=
  \frac{\partial_\lambda \theta \ \partial_\theta \mathcal{S}_A}
       {\partial_\lambda \theta \ \partial_\theta \mathcal{E}_A}
       =-\frac{1}{\epsilon_A} \ln \tan^2{\theta}
       =\beta_A^\star.
       \label{BCooper}
\end{equation}
From Eqs~(\ref{rhoredCooper})--(\ref{betaCooper}) and (\ref{BCooper}), we can conclude that 
the relation 
\begin{equation}
  \hat{\rho}_{A}^{\rm can}(\mathcal{B}_A(\lambda)) = \hat{\rho}_{A}^{\rm red}(\lambda)
  \label{Fcan_Fred}
\end{equation}
holds exactly and thus is no longer a conjecture in this case.

We note that the relation in Eq.~(\ref{BCooper}) can be
derived also by directly
calculating $\partial \mathcal{S}_{A}/\partial \mathcal{E}_{A}$. 
Namely, it follows from Eq.~(\ref{ECooper}) that 
$\sin^2{\theta}=\frac{1}{2}+\frac{\mathcal{E}_A}{\epsilon_A}$ and 
$\cos^2{\theta}=\frac{1}{2}-\frac{\mathcal{E}_A}{\epsilon_A}$. 
Therefore, $\mathcal{S}_{A}$ can be expressed in terms of $\mathcal{E}_{A}$ as
\begin{equation}
  \mathcal{S}_{A} =
  -
  \left(\frac{1}{2}-\frac{\mathcal{E}_A}{\epsilon_A}\right) \ln
  \left(\frac{1}{2}-\frac{\mathcal{E}_A}{\epsilon_A}\right)
  -
  \left(\frac{1}{2}+\frac{\mathcal{E}_A}{\epsilon_A}\right) \ln
  \left(\frac{1}{2}+\frac{\mathcal{E}_A}{\epsilon_A}\right),  
\end{equation}
and thus one can readily show that  
\begin{equation}
  \frac{\partial \mathcal{S}_A}{\partial \mathcal{E}_A}
  =\beta_A^\star.  
\end{equation}

To confirm more specifically the correspondence
between the two density matrix operators in
Eq.~(\ref{Fcan_Fred}),  
let us rewrite
the entanglement von Neumann entropy $\mathcal{S}_A$ and
the energy $\mathcal{E}_A$
in terms of $\beta_A^\star$, instead of $\theta$. 
It follows from Eq.~(\ref{betaCooper}) that
\begin{alignat}{1}
  &\tan^2{\theta}=\e^{-\beta_A^\star \epsilon_A},\\
  &\cos^2\theta=1-f_{\beta_A^\star}(\epsilon_A), \label{cos2}
\end{alignat}
and
\begin{alignat}{1}
  &\sin^2\theta=f_{\beta_A^\star}(\epsilon_A),     \label{sin2}
\end{alignat}
where
\begin{equation}
  f_{\beta_A^\star}(\epsilon_A)= \frac{1}{\e^{\beta_A^\star \epsilon_A}+1} 
\end{equation}
is the Fermi-Dirac distribution function at inverse temperature $\beta_A^\star$.
Substituting Eqs.~(\ref{cos2}) and (\ref{sin2})
into Eqs.~(\ref{SCooper}) and (\ref{ECooper}) yields that
\begin{alignat}{1}
  \mathcal{S}_{A} &= -(1-f_{\beta_A^\star})
  \ln(1-f_{\beta_A^\star})
  -f_{\beta_A^\star} \ln f_{\beta_A^\star} 
\end{alignat}
and
\begin{alignat}{1}
  \mathcal{E}_{A} &= \epsilon_A \left(f_{\beta_A^\star}-\frac{1}{2}\right), 
\end{alignat}
which are familiar forms of the thermodynamic 
entropy and the internal energy of
free fermions, respectively. 
One can also readily find that
the positive square root $Z_A
=\e^{\beta^\star_A\epsilon_A/2}+\e^{-\beta^\star_A\epsilon_A/2}>0$ of $Z_A^2$ 
gives the corresponding partition function, i.e.,
$Z_A={\rm Tr}_{A}\left[ \e^{-\beta^\star_A \hat{H}_A} \right]$.

By doing the same analysis for subsystem $B$, 
one can find the relation between the effective temperatures
$\mathcal{T}_A(\lambda)=\mathcal{B}_A^{-1}(\lambda)$ and 
$\mathcal{T}_B(\lambda)=\mathcal{B}_B^{-1}(\lambda)$ as 
\begin{equation}
  \mathcal{T}_{A}(\lambda)/\epsilon_A=\mathcal{T}_{B}(\lambda)/\epsilon_B.
  \label{eq:temp_F}
\end{equation}
We also note that all these analyses given above are based on the ground state
in Eq.~(\ref{GSCooper}) under the conditions in Eq.~(\ref{Fstable}). 
As in the bosonic case, 
one can readily show that if $\epsilon_A\not = \epsilon_B$, 
the maximum of the effective temperature is bounded. 
For example, when 
$\epsilon_B/\epsilon_A<1$, 
$\tan^2{\theta}<\epsilon_B/\epsilon_A$ should be satisfied 
in order to satisfy the conditions in Eq.~(\ref{Fstable}). 
This implies that 
$ \ln{(\epsilon_A/\epsilon_B)} < \beta_A^\star \epsilon_A<\infty$,
or equivalently 
$ 0 < T_A^\star/\epsilon_A< 1/\ln{(\epsilon_A/\epsilon_B)}$.

Finally, we briefly describe the correspondence between the BCS Hamiltonian 
and the present Hamiltonian discussed in this section. 
The BCS Hamiltonian ${\hat H}_{\rm BCS}$ is generally described by the following 
Hamiltonian in the momentum space: 
\begin{equation}
{\hat H}_{\rm BCS}  =  \sum_{{\bm k},\sigma} \xi_{\bm k} {\hat c}_{{\bm k}\sigma}^\dag {\hat c}_{{\bm k}\sigma} 
+
\sum_{\bm k}
\Delta_{\bm k}
\left(
{\hat c}_{{\bm k}\uparrow} {\hat c}_{-{\bm k}\downarrow} 
+
{\hat c}_{-{\bm k}\downarrow}^\dag {\hat c}_{{\bm k}\uparrow}^\dag \right),
\end{equation}
where ${\hat c}_{{\bm k}\sigma}^\dag$ (${\hat c}_{{\bm k}\sigma}$) is the creation (annihilation) operator of electron 
with momentum $\bm k$ and spin $\sigma\,(=\uparrow,\downarrow)$, $\xi_{\bm k}=\varepsilon_{\bm k}-\mu$, 
$\varepsilon_{\bm k}$ is the single-particle energy dispersion of electrons, $\mu$ is the chemical potential, 
and $\Delta_{\bm k}$ is the gap function and is assumed to be real.
Note that the spatial dimensionality is not assumed. 
We now introduce the following canonical transformation: 
\begin{alignat}{3}
  &{\hat c}_{{\bm k}\uparrow}^\dag \to {\hat a}_{\bm k}^\dag,
  &{\hat c}_{-{\bm k}\downarrow}^\dag \to {\hat b}_{\bm k}^\dag \quad  
  &({\rm for}\,\, \,\xi_{\bm k}>0), \\
  &{\hat c}_{-{\bm k}\downarrow} \to {\hat a}_{\bm k}^\dag,
  &{\hat c}_{{\bm k}\uparrow} \to {\hat b}_{\bm k}^\dag \quad
  &({\rm for}\,\, \,\xi_{\bm k} \leqslant 0), 
\end{alignat}
where ${\hat a}_{\bm k}^\dag$ and ${\hat a}_{\bm k}$ (${\hat b}_{\bm k}^\dag$ and ${\hat b}_{\bm k}$) satisfy the anticommutation 
relations, e.g., $\{{\hat a}_{\bm k}, {\hat a}_{{\bm k}'}^\dag\}=\delta_{{\bm k},{\bm k}'}$ and $\{{\hat a}_{\bm k}, {\hat b}_{{\bm k}'}^\dag\}=0$. 
With this canonical transformation, the BCS Hamiltonian is rewritten as 
\begin{alignat}{1}
{\hat H}_{\rm BCS} & = 
\sum_{{\bm k}\,(\xi_{\bm k}>0)}\left[ \xi_{\bm k}\left( {\hat a}_{\bm k}^\dag{\hat a}_{\bm k} + {\hat b}_{\bm k}^\dag{\hat b}_{\bm k} \right) 
  +
  \Delta_{\bm k}
  \left(
        {\hat a}_{\bm k} {\hat b}_{\bm k} +
        {\hat b}_{\bm k}^\dag {\hat a}_{\bm k}^\dag \right)  \right] \nonumber \\
&+ \sum_{{\bm k}\,(\xi_{\bm k} \leqslant 0)}\left[ -\xi_{\bm k}\left( {\hat a}_{\bm k}^\dag{\hat a}_{\bm k} + {\hat b}_{\bm k}^\dag{\hat b}_{\bm k} \right) 
  +
  \Delta_{\bm k}
  \left(
       {\hat a}_{\bm k} {\hat b}_{\bm k} +
       {\hat b}_{\bm k}^\dag {\hat a}_{\bm k}^\dag \right)  \right] \nonumber \\
&+ 2 \sum_{{\bm k}\,(\xi_{\bm k} \leqslant 0)} \xi_{\bm k}
\\
&= \sum_{{\bm k}}\left[ \left|\xi_{\bm k}\right|\left( {\hat a}_{\bm k}^\dag{\hat a}_{\bm k} -\frac{1}{2}\right) 
+ \left|\xi_{\bm k}\right|\left( {\hat b}_{\bm k}^\dag{\hat b}_{\bm k}-\frac{1}{2} \right) \right. \nonumber \\
& \quad\quad\quad\quad\quad\quad +\Delta_{\bm k} \left(  {\hat a}_{\bm k} {\hat b}_{\bm k} + {\hat b}_{\bm k}^\dag {\hat a}_{\bm k}^\dag \right)  \Biggr] 
  + \sum_{\bm k} \xi_{\bm k}.
\end{alignat}
Here $\sum_{{\bm k}\,(\xi_{\bm k}>0)}$ ($\sum_{{\bm k}\,(\xi_{\bm k} \leqslant  0)}$) indicates the sum over $\bm k$ 
with $\xi_{\bm k}>0$ ($\xi_{\bm k} \leqslant 0$) and 
we assume that $\xi_{\bm k} = \xi_{-\bm k}$. 
Therefore, apart from the irrelevant constant term, each component with a given momentum $\bm k$ in the BCS Hamiltonian 
${\hat H}_{\rm BCS}$ is exactly the same as the Hamiltonian $\hat H(\lambda)=\hat H_A + \hat H_B + \hat V_{AB}(\lambda)$ 
in Eqs~(\ref{FHA})--(\ref{FVAB}) with the correspondence of
$\epsilon_A=\epsilon_B\leftrightarrow |\xi_{\bm k}|$ and 
$\lambda\leftrightarrow\Delta_{\bm k}$. 
For example, the ground state of $\hat H_{\rm BCS}$ is thus given simply as a product state of $|\Psi_0(\lambda=\Delta_{\bm k})\rangle$ 
in Eq.~(\ref{GSCooper}) over all momenta. 
Following the same argument given above in this section, we can conclude that the reduced density matrix operator of subsystem $A$ 
for the ground state of the BCS Hamiltonian is exactly the same as the thermodynamic density matrix operator of the isolated 
subsystem $A$ with the effective temperature introduced in Eq.~(\ref{beta_dSdE}). However, we should note that bipartitioning  
of the whole Hilbert space is not trivial because the subsystem $A$ consists of Hilbert space for up electrons with $\xi_{\bm k}>0$ 
and down electrons with $\xi_{\bm k} \leqslant 0$, i.e., the subsystem $A$ being described by 
\begin{alignat}{1}
  \hat H_A &=
  \sum_{{\bm k}}           |\xi_{\bm k}|\left({\hat a}_{{\bm k}}^\dag {\hat a}_{{\bm k}} -\frac{1}{2}\right)
    \notag \\
&= \sum_{{\bm k}\,(\xi_{\bm k}>0)} \xi_{\bm k} \left({\hat c}_{{\bm k}\uparrow}^\dag {\hat c}_{{\bm k}\uparrow} -\frac{1}{2}\right)
+ \sum_{{\bm k}\,(\xi_{\bm k} \leqslant  0)} \xi_{\bm k} \left({\hat c}_{{\bm k}\downarrow}^\dag {\hat c}_{{\bm k}\downarrow} - \frac{1}{2}\right). 
\end{alignat}

\section{Discussions}\label{sec.discussions}

\subsection{Insights of the effective inverse temperature $\mathcal{B}_A(\lambda) $}\label{sec:beta}
The observation from the numerical calculations in Sec.~\ref{sec.results} as well as two analytical examples 
in Sec.~\ref{sec:analytical} 
leads us to conjecture that a canonical ensemble with the inverse temperature 
\begin{equation}
  \beta =\mathcal{B}_A(\lambda)     
  \label{beta_beta}
\end{equation}
could emerge by quantum entanglement in a partitioned subsystem of a pure ground state. 
This assertion is highly nontrivial as $\beta$
in the left-hand side is a given 
inverse temperature in the canonical ensemble,
while $\mathcal{B}_A(\lambda)$ in the 
right-hand side is evaluated 
in the entangled pure ground state $|\Psi_0(\lambda)\rangle$
of $\hat{H}(\lambda)$.
Here, we further discuss 
the observation summarized in Eqs.~(\ref{eq.rhorho}) and (\ref{beta_beta})
to gain more insights. 
Note however that
we do not intend to prove
Eq.~(\ref{eq.rhorho}) or (\ref{beta_beta}).

\subsubsection{Product state and additivity}
Let us first briefly review the additivity of
the entanglement von Neumann entropy by considering a product state~\cite{Wehrl1978}. 
Consider a system $W$ that is composed of subsystems $X$ and $Y$.  
Note that these are nothing to do with system $A+B$ consisting of 
subsystems $A$ and $B$ considered in the previous sections. 
Let $\hat{\rho}_{X\,(Y)}$ be the density matrix operator of subsystem $X\,(Y)$, 
and suppose that the density matrix operator $\hat{\rho}_{W}$ of the total system 
is given as a product state of $\hat{\rho}_{X}$ and $\hat{\rho}_{Y}$, i.e., 
\begin{equation}
  \hat{\rho}_{W} = \hat{\rho}_{X} \otimes \hat{\rho}_{Y}, 
  \label{rhopower}
\end{equation}
implying no entanglement between subsystems $X$ and $Y$.
Then the entanglement von Neumann entropy $\mathcal{S}_s$ 
($s=W, X,$ or $Y$) defined as
\begin{equation}
  \mathcal{S}_{s} = - {\rm Tr}_s \left[\hat{\rho}_s \ln \hat{\rho}_s \right]
\end{equation}
possesses the additivity
\begin{equation}
  \mathcal{S}_{W}=\mathcal{S}_X + \mathcal{S}_Y. 
\end{equation}

Next, let us consider the additivity of the energy. 
For this purpose, we introduce Hamiltonian. 
Let $\hat{H}_{X\,(Y)}$ be the Hamiltonian of subsystem $X\,(Y)$, 
and suppose that the total Hamiltonian $\hat{H}_{W}$ of the system $W$
is given as
\begin{equation}
  \hat{H}_{W} =
  \hat{H}_{X} \otimes \hat{I}_{Y} + 
  \hat{I}_{X} \otimes \hat{H}_{Y}, 
\end{equation}
implying no interaction between subsystems $X$ and $Y$.
Notice that any eigenstate of $\hat{H}_{W}$ is 
given as a product of eigenstates of $\hat{H}_{X}$ and $\hat{H}_Y$, 
satisfying the form in Eq.~(\ref{rhopower}). 
Then the energy $\mathcal{E}_s$ defined as 
\begin{equation}
  \mathcal{E}_{s} = {\rm Tr}_s \left[\hat{\rho}_s \hat{H}_s \right] 
\end{equation}
possesses the additivity 
\begin{equation}
  \mathcal{E}_{W} =\mathcal{E}_X + \mathcal{E}_Y. 
\end{equation}

\subsubsection{Functional form of density matrix operator}\label{sec.assumption}
Now we show that the Gibbs state, i.e., the thermodynamic density matrix operator, arises if 
a particular functional form for a density matrix operators is assumed. 
Let us assume that 
the density matrix operator $\hat{\rho}_{s}$ of each system
depends on its own Hamiltonian $\hat{H}_s$ with 
a common functional form $\rho(\cdot)$, i.e., 
\begin{equation}
  \hat{\rho}_{s} \overset{!}{=} \rho({\hat{H}_s}). 
  \label{rhoH}
\end{equation}
This assumption implies that $\hat{\rho}_s$ commutes with  $\hat{H}_{s}$
and hence $\hat{\rho}_s$ and $\hat{H}_s$ are simultaneously diagonalizable.
According to the Liouville-von Neumann equation   
$\imag \frac{\partial \hat{\rho}_s }{\partial t}
= [\hat{H}_s,\hat{\rho}_s]$ with $t$ being the time, 
$\hat{\rho}_s$ in the form of Eq.~(\ref{rhoH}) 
is a stationary state that does not evolve in time
via the unitary evolution with the Hamiltonian $\hat{H}_s$.

Under the assumption in Eq.~(\ref{rhoH}),
Eq.~(\ref{rhopower}) can be written as 
\begin{equation}
  \rho(
  \hat{H}_X \otimes \hat{I}_Y +
  \hat{I}_X \otimes \hat{H}_Y )
  =
  \rho(\hat{H}_X) \otimes
  \rho(\hat{H}_Y).
  \label{rhopowerlaw}
\end{equation}
Equation~(\ref{rhopowerlaw}) implies that $\rho(\hat{H}_s)$
is an exponential function of $\hat{H}_s$.  
Further taking into account
the Hermiticity $\rho(\hat{H}_s)^\dag =\rho(\hat{H}_s)$ and
the normalization ${\rm Tr}_s[\rho(\hat{H}_s)]=1$, we can infer that 
the form of $\rho(\hat{H}_s)$ should be  
\begin{equation}
  \rho({\hat{H}_s}) = \frac{\e^{-\beta^\star \hat{H}_s}}{{\rm Tr}_s[\e^{-\beta^\star \hat{H}_s}] }
  = \hat{\rho}_{s}^{\rm can}(\beta^\star)
  \label{rhoexpH}
\end{equation}
with $\beta^\star$ real. 
Note that $\beta^\star$ can be either negative or positive,  
provided that the spectrum of $\hat{H}_s$ is bounded.
Obviously from the assumption in Eq.~(\ref{rhoH}),
$\beta^\star$ is common in subsystems $X$ and $Y$ as well as the system $W$, 
otherwise Eq.~(\ref{rhoexpH}) does not satisfy Eq.~(\ref{rhopowerlaw})
in general. 
Such a ``common temperature'' property of $\beta^\star$
required for the additivity of 
the entanglement von Neumann entropy and the energy is analogous to
the property of the thermodynamic temperature 
characterizing equilibrium between subsystems $X$ and $Y$. 
Thus the Gibbs state as well as
the inverse-temperature-like real number $\beta^\star$ 
have arisen from the assumption in Eq.~(\ref{rhoH}).

\subsubsection{$\beta^\star$ as a derivative of entanglement entropy and energy}\label{sec.motive}
Now we derive Eq.~(\ref{beta_beta}) by assuming the functional form of Eq.~(\ref{rhoH}) 
even when there exists an interaction between subsystems, as in the case studied in Sec.~\ref{sec.results}.  
Under this assumption, the reduced density matrix operator $\hat{\rho}_A^{\rm red}(\lambda)$ of subsystem $A$ has
the form as in Eq.~(\ref{rhoexpH}), i.e., 
\begin{equation}
  \hat{\rho}_A^{\rm red}(\lambda) \overset{!}{=}
  \hat{\rho}_A^{\rm can}(\beta^\star) =
  \frac{\e^{-\beta^\star \hat{H}_A}}{Z_A(\beta^\star)}
  \label{can}
\end{equation}
with $\beta^\star$ real. 
In our setting, the parameter $\lambda$ does not enter in 
$\hat{H}_A$ but 
$\hat{\rho}_{A}^{\rm red}(\lambda)$ 
should depends on $\lambda$ through $\beta^\star$, i.e.,
\begin{equation}
  \beta^\star = \beta^\star(\lambda).
\end{equation}
As described in details in Appendix~\ref{app}, considering the relative entropy
$D(\hat{\rho}_1|\hat{\rho}_0)
=
{\rm Tr}\left[
  \hat{\rho}_1 \ln \hat{\rho}_1
  \right]
-
{\rm Tr}\left[
  \hat{\rho}_1 \ln \hat{\rho}_0
  \right]$ 
with 
$\hat{\rho}_0 = \hat{\rho}_{A}^{\rm red}(\lambda)$ and 
$\hat{\rho}_1 = \hat{\rho}_{A}^{\rm red}(\lambda+\Delta \lambda)$,
we obtain that 
\begin{eqnarray}
  D(\hat{\rho}_1|\hat{\rho}_0)
  &=&
  \beta^\star(\lambda)
  \left[\mathcal{E}_A(\lambda+\Delta \lambda)-\mathcal{E}_A(\lambda)\right]\notag \\
  &-&\left[\mathcal{S}_A(\lambda+\Delta \lambda)-\mathcal{S}_A(\lambda)\right].
\end{eqnarray}
Since $D(\hat{\rho}_1|\hat{\rho}_0)=O((\Delta\lambda)^2)$ 
[see Ref.~\cite{Blanco2013} and also Eq.~(\ref{eq.Dvanish}) in Appendix~\ref{app}], 
we finally obtain, by solving the above equation with respect to $\beta^\star(\lambda)$, that  
\begin{eqnarray}
  \beta^\star(\lambda)
  &=& \frac
  {\mathcal{S}_A(\lambda+\Delta \lambda) - \mathcal{S}_A(\lambda)}
  {\mathcal{E}_A(\lambda+\Delta \lambda) - \mathcal{E}_A(\lambda)}
  + O\left((\Delta \lambda)^2 \right) \notag \\
  &\underset{\Delta \lambda \to 0}{=}& \mathcal{B}_A(\lambda), \label{eq:beta_a}
\end{eqnarray}
leading to the form of Eq.~(\ref{beta_dSdE}) and 
consistent with 
the observation in Eqs.~(\ref{eq.rhorho}) and (\ref{beta_beta}).

Remarkably, the functional form of the reduced density matrix operator 
as in Eq.~(\ref{rhoH}) has been proven to be valid for
a certain class of topological quantum states~\cite{Qi2012} and we have also already 
shown that it is the case for the two examples described in Sec.~\ref{sec:analytical}. 
Although such a dependence of the reduced density matrix operator
on the Hamiltonian is in general not necessarily valid, 
our numerical results 
suggest that the reduced density matrix operator of a partitioned subsystem for the ground state 
of the Heisenberg models in the two-leg ladder and the bilayer lattices
can be well approximated in the form of Eq.~(\ref{rhoH}), 
which describes the Gibbs state with 
the effective inverse temperature $\mathcal{B}_A(\lambda)$.  
Finally, we note that
the effective inverse temperature $\mathcal{B}_A(\lambda)$ of subsystem $A$ differs in general from
the effective inverse temperature $\mathcal{B}_B(\lambda)$ of subsystem $B$ 
(see Sec.~\ref{sec:analytical} and Appendix~\ref{app.TB}).

\subsection{Thermal and quantum fluctuations}
Let us now discuss an association between thermal and quantum fluctuations. 
The quantities 
$W_A\equiv\e^{S_A}$ and $\mathcal{W}_A\equiv\e^{\mathcal{S}_A}$,
each satisfying 
$1 \leqslant W_A \leqslant D_A$ and 
$1 \leqslant \mathcal{W}_A \leqslant D_A$, 
can be regarded as effective numbers of microscopic pure states that 
contribute to the thermodynamic and reduced density matrix operators,
respectively.
Considering that fluctuations are induced by
a statistical mixture of microscopic states
in a density matrix operator, 
$S_{A}$ and $\mathcal{S}_A$ 
may serve as a measure 
of the thermal fluctuation due to the temperature and 
as a measure of the quantum fluctuation due to the quantum entanglement, respectively. 
The almost indistinguishable agreement between 
$S_A$ versus $T$ and $\mathcal{S}_A$ versus $\mathcal{T}_A(\lambda)$ found numerically in Sec.~\ref{sec.results} 
(and also the exact agreement in the case of two analytical examples in Sec.~\ref{sec:analytical})  
suggests that the quantum fluctuation
in the partitioned subsystem $A$ coupled to the other subsystem via the coupling parameter $\lambda$
can mimic the thermal fluctuation
in the isolated subsystem $A$ at the temperature $T=\mathcal{T}_A(\lambda)$
and vice versa (see Fig.~\ref{fig.schematic}).
In other words, the mixture of microscopic states caused by  
either temperature or quantum entanglement 
is essentially indistinguishable, at least, for 
the quantities studied here. 
A related discussion on thermal and quantum
fluctuations has also been reported in Ref.~\cite{Sugiura2014}.

\subsection{Ground-state degeneracy}

Our numerical results involving the reduced density matrix operator in Sec.~\ref{sec.results}
are obtained for the unique ground state $|\Psi_0\rangle$. 
%for all the cases.  
Generally, the ground state of a finite-size system is 
unique and does not break any symmetry~\cite{Shimizu2001,Shimizu2002}.
However, either by a careful choice of a model or by an accident,
the ground state could be degenerate even in a finite-size system. 
Here we shall briefly explain that an ambiguity occurs for 
determining the reduced density matrix operator 
when the ground state of the whole system is degenerate. 

Suppose that the ground state of the whole system is $g$-fold
degenerate with the degenerate ground states 
$\{|\Psi_d\rangle\}_{d=0}^{g-1}$,  
each satisfying $\hat{H}|\Psi_d\rangle = E_0 |\Psi_d\rangle$. 
Without loss of generality, 
$\{|\Psi_d\rangle\}_{d=0}^{g-1}$
can be chosen to satisfy 
$\langle\Psi_d|\Psi_{d^\prime}\rangle=\delta_{dd^\prime}$
by using, e.g.,  a Gram-Schmidt orthonormalization method.
Then, any linear combination of these states, 
$|{\rm GS}\rangle \equiv \sum_{d=0}^{g-1} \alpha_{d} |\Psi_d\rangle$ 
with $\sum_{d=0}^{g-1}|\alpha_d|^2=1$, is a normalized pure ground state
because $\hat{H}|{\rm GS}\rangle = E_0 |{\rm GS}\rangle$. 
Apparently, the reduced density matrix operator, 
$\hat{\rho}^{\rm red} = {\rm Tr}_B [|{\rm GS}\rangle \langle {\rm GS}|]$,  
depends on the choice of the coefficients $\{\alpha_d\}_{d=0}^{g-1}$
and hence is not uniquely determined.

If we define the ground state as a state 
for which the expectation value of $\hat{H}$ is $E_0$, 
then the ground state with $g \geqslant 2$ can be 
represented also as a mixed state of the form 
$\hat{\rho}_{\rm GS}\equiv \sum_{d=0}^{g-1}p_d |\Psi_d\rangle \langle \Psi_d|$
with $\sum_{d=0}^{g-1}p_d=1$ and $p_d \geqslant 0$.
The mixed ground state $\hat{\rho}_{\rm GS}$ can be considered as a linear combination of 
the projectors, $\{|\Psi_d\rangle \langle \Psi_d|\}_{d=0}^{g-1}$, onto the
eigenspace of the ground state. 
Indeed, $\hat{\rho}_{\rm GS}$ satisfies 
${\rm Tr}[\hat{\rho}_{\rm GS}]=1$, 
$\hat{\rho}_{\rm GS} \hat{H}=\hat{H}\hat{\rho}_{\rm GS}=E_0\hat{\rho}_{\rm GS}$, and hence 
${\rm Tr}[\hat{\rho}_{\rm GS} \hat{H}]=E_0$. 
It is also apparent that 
the reduced density matrix operator,
$\hat{\rho}^{\rm red}={\rm Tr}_{B}[\hat{\rho}_{\rm GS}]$, 
is not uniquely determined because it 
depends on the choice of the coefficients $\{p_d\}_{d=0}^{g-1}$. 

In either case, the degeneracy in the ground state of
the whole system leads to an ambiguity for determining the reduced density matrix operator,
as the reduced density matrix operator 
depends on the choice of the degenerate ground state of the whole system. 
To avoid such an ambiguity, it is crucial that
the ground state of the whole system is unique.

\section{Conclusion and remarks}\label{sec.conclusions}
In conclusion, by numerically analyzing the spin-1/2 antiferromagnetic Heisenberg model 
in the two-leg ladder and the bilayer lattices, we have 
examined the emergence of a thermal equilibrium 
in a partitioned subsystem $A$ of a pure ground state by quantum entanglement. 
Under the bipartitioning of the whole system into subsystems with the
entanglement cut that covers the entire volume of subsystem $A$, our numerical calculations 
for the entanglement von Neumann entropy and the energy of subsystem $A$ strongly support
the emergent thermal equilibrium 
that numerically agrees well with
the canonical ensemble, 
where  
the temperature $\mathcal T_A(\lambda)$ in the canonical ensemble is 
determined from the entanglement von Neumann entropy and the energy 
of subsystem $A$. 
The fidelity calculations ascertain
that the reduced density matrix operator of subsystem $A$ 
matches, within the maximum error of $1.5\%$ in the finite size clusters studied, 
the Gibbs state, i.e., thermodynamic density matrix operator, with temperature $\mathcal T_A(\lambda)$. 
Furthermore, we have found that, apart from the case of the bilayer triangular lattice with the smallest cluster, 
the temperature $\mathcal T_A(\lambda)$ calculated from the ground state of the whole system
depends insignificantly 
on the system sizes, in good accordance with the fact the thermodynamic temperature is an intensive quantity.
Our numerical finding is further supported by two simple but nontrivial examples, for which one can show analytically that 
the two density matrix operators are exactly the same with temperature $\mathcal T_A(\lambda)$.

Once we accept that the reduced density matrix operator $\hat{\rho}_{A}^{\rm red}(\lambda)$ represents 
a thermodynamic density matrix operator that describes a statistical ensemble 
of subsystem $A$ at thermodynamic temperature
$\mathcal{T}_A(\lambda)$, 
our scheme provides an alternative way to calculate finite-temperature
properties based on pure ground-state quantum-mechanical calculations, 
as demonstrated in Sec.~\ref{sec.generalT}. 
Our scheme is similar to those based on 
purification~\cite{Suzuki1985,Verstraete2004,Zwolak2004,Feiguin2005,Wu2019}
(see Appendix~\ref{app.TFD}), but has several advantages. For example, 
a parallel calculation with respect to different temperatures is possible merely by calculating the ground states 
$|\Psi_0(\lambda)\rangle$ with different values of $\lambda$ independently, and
no imaginary-time-evolution-type
calculations, which apply $\exp(-\beta \hat{H}_A) \otimes \hat{I}_B$ to some states, are required.
However, it is not straightforward to have a desired ``temperature'' 
because $\mathcal{T}_A(\lambda)$ is not an input parameter
but is evaluated from the entanglement von Neumann entropy and the energy, 
similarly to microcanonical ensemble methods~\cite{Long2003,Sugiura2012,Okamoto2018}.

In order to have quantitative agreement between   
the entanglement von Neumann entropy $\mathcal{S}_A$ and  
the thermodynamic entropy $S_A$, 
the entanglement von Neumann entropy $\mathcal{S}_{A}$ 
should obey the volume law, instead of the area law, because the thermodynamic entropy $S_A$ is an extensive quantity. 
This implies that the entanglement cut
for bipartitioning the system should cover the entire volume of the subsystem, 
as we have considered in this study.
This also implies that 
the subsystem $B$ should be at least as large as the subsystem $A$, 
i.e., $N_{B} \geqslant N_{A}$. 
The lower bound $N_{B}=N_{A}$, or equivalently $N=N_{A}+N_{B}=2N_A$,  
is in fact consistent with the system size that is required for 
the purification of a mixed state $\hat{\rho}_A^{\rm red}$.

Technically,
the doubling of the system size $N_A$
for a pure ground state calculation 
becomes immediately intractable with increasing $N_A$ by 
the exact diagonalization method simply 
because of the exponential increase of the computational cost
with respect to the system size.
The density matrix renormalization group (DMRG) method~\cite{White1992} 
might be a choice of methods for overcoming this difficulty
especially for 1D systems. 
However, since the entanglement von Neumann entropy should obey the volume law,  
a large amount of computational resource may be required  
in DMRG calculations to obtain accurate results even
for 1D systems. Another possibility would be 
quantum computation for many-body systems, 
for which a quantum algorithm to compute the
entanglement spectrum~\cite{Johri2017} can be used.

Although the fidelity $F$ of the two density matrix operators is found to be close to 1, 
the largest deviation from 1 occurs at some particular $\mathcal T_A(\lambda)$, around which the finite size effect seems to be 
the largest (see Fig.~\ref{fig.fidelity} and Fig.~\ref{fig.correlation}). 
Therefore, it is desirable to examine the finite size effect more systematically.  
We have considered only three particular systems numerically.
The extension of the present study to other systems
such as larger spins or interacting fermionic systems is also highly interesting 
to understand under what conditions
a thermal equilibrium can emerge in a subsystem 
of a pure ground state by quantum entanglement. 
These studies are certainly beyond the currently available computational power and are left for future work. 
Finally, we note that for testing and extending the present study, 
not only numerical calculations, 
but rather experiments for cold atom systems~\cite{Islam2015,Kaufman2016}
would be promising.

\acknowledgments
The authors would like to thank
Tomonori Shirakawa and Hiroaki Matsueda 
for valuable discussions.
Parts of numerical simulations have been done
on the HOKUSAI supercomputer at RIKEN 
(Project ID: G20015).
This work was supported by Grant-in-Aid
for Research Activity start-up (No.~JP19K23433) and 
Grant-in-Aid for Scientific Research (B) (No.~JP18H01183) from MEXT, Japan.

\appendix
\section{Relative entropy}~\label{app}
\subsection{Definition}
For density matrix operators $\hat{\rho}_0$ and $\hat{\rho}_1$,
the relative entropy $D(\hat{\rho}_1|\hat{\rho}_0)$ is defined as~\cite{Umegaki1962,Lindblad1973,Sagawa2012}
\begin{equation}
  D(\hat{\rho}_1|\hat{\rho}_0)
  =
  {\rm Tr}\left[
    \hat{\rho}_1 \ln \hat{\rho}_1
    \right]
  -
  {\rm Tr}\left[
    \hat{\rho}_1 \ln \hat{\rho}_0
    \right]
  \geqslant 0.
\end{equation}
The equality is satisfied if and only if $\hat{\rho}_1=\hat{\rho}_0$. 
In terms of the entanglement entropy 
$\mathcal{S}(\hat{\rho})=-{\rm Tr}[\hat{\rho} \ln \hat{\rho}]$, 
the relative entropy can be rewritten as
\begin{eqnarray}
  D(\hat{\rho}_1|\hat{\rho}_0)
  =
  -{\rm Tr}\left[ (\hat{\rho}_1-\hat{\rho}_0) \ln \hat{\rho}_0 \right]
  -\left[\mathcal{S}(\hat{\rho}_1)-\mathcal{S}(\hat{\rho}_0)\right]. 
  \label{D10}
\end{eqnarray}

\subsection{Relative entropy for two close density matrix operators}
\begin{figure}
  \begin{center}
    \includegraphics[width=0.75\columnwidth]{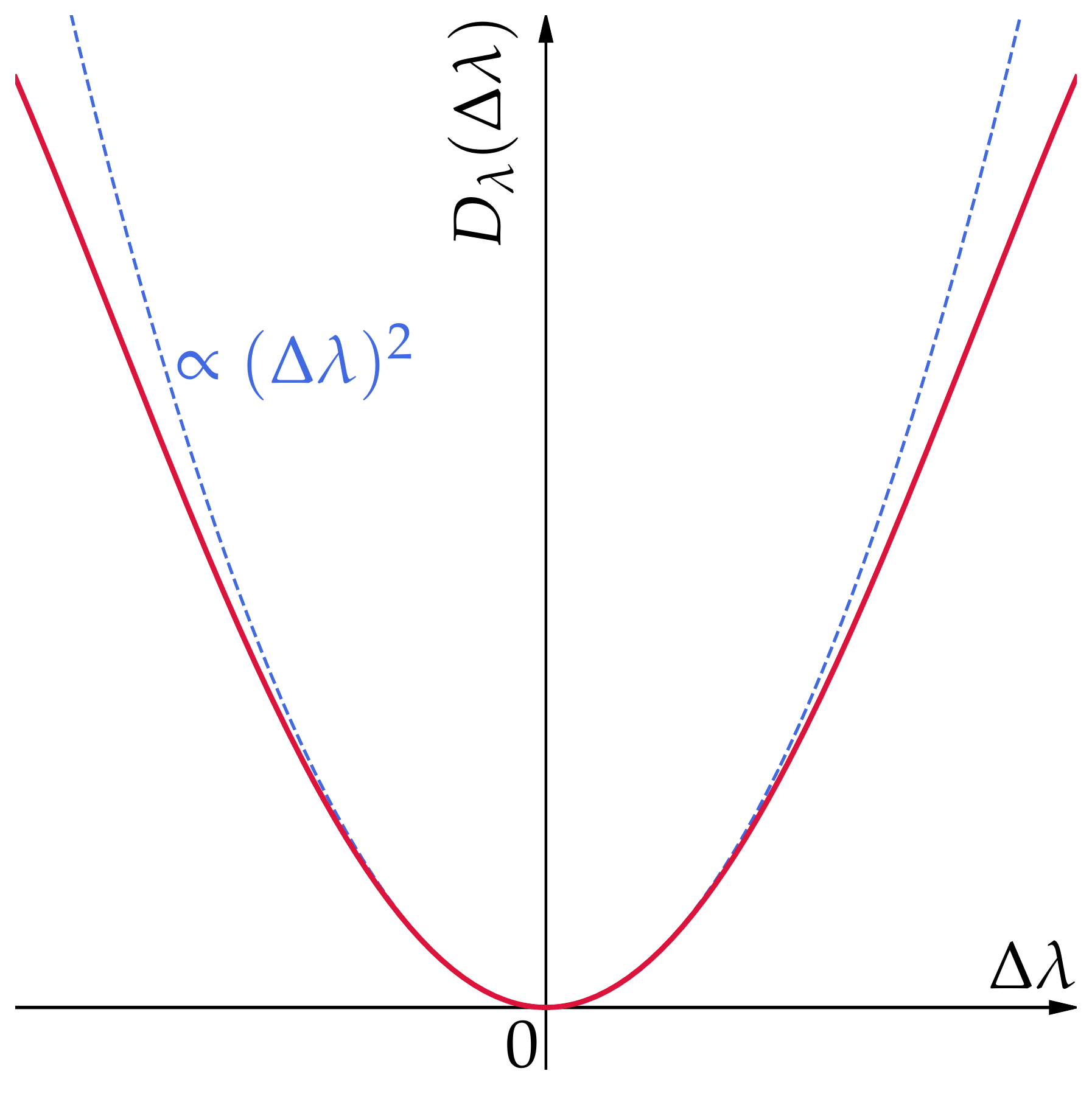}
    \caption{
      Schematic figure of the relative entropy
      $D_{\lambda}(\Delta \lambda) = D(\hat{\rho}(\lambda+\Delta \lambda)|\hat{\rho}(\lambda))$ for small $|\Delta \lambda|$ 
      (red thick line),
      showing that $D_{\lambda}(0)=0$ and $\dd D_{\lambda}(0)/\dd \lambda=0$. Blue dashed line indicates a function proportional 
      to $(\Delta\lambda)^2$, for comparison. 
      \label{Dlambda}}    
  \end{center}
\end{figure}

We now consider a density matrix operator
$\hat{\rho}(\lambda)$ parametrized by $\lambda$. 
Suppose that
$\hat{\rho}_0=\hat{\rho}(\lambda)$ and
$\hat{\rho}_1=\hat{\rho}(\lambda+\Delta \lambda)$. 
For convenience, let us simply write the relative entropy as 
\begin{equation}
  D_{\lambda}(\Delta \lambda)
  \equiv D(\hat{\rho}(\lambda+\Delta \lambda)|\hat{\rho}(\lambda)). 
\end{equation}
The Taylor expansion of $D_{\lambda}(\Delta \lambda)$
around  $\Delta \lambda = 0$ is 
\begin{equation}
  D_{\lambda}(\Delta \lambda)
  =
  D_{\lambda}(0)+
  \left.
  \frac{\dd D_{\lambda}(\Delta \lambda)}
       {\dd \lambda}        \right|_{\Delta \lambda=0}
       \Delta \lambda       
       +O\left( (\Delta \lambda)^2 \right).
       \label{D_Taylor}
\end{equation}
Since $D_{\lambda}(0)=0$, the first term in the right hand side of Eq.~(\ref{D_Taylor}) vanishes. 
Moreover, the first derivative vanishes at $\Delta \lambda =0$, i.e., 
\begin{equation}
  \left.
  \frac{\dd D_{\lambda}(\Delta \lambda)}
       {\dd \lambda}        \right|_{\Delta \lambda=0}
       = 0.
       \label{D_1vanish}
\end{equation}
This can be shown by substituting the Taylor expansion 
$\hat{\rho}(\lambda+\Delta \lambda) = \hat{\rho}(\lambda)
+ \frac{\dd \hat{\rho}(\lambda)}{\dd \lambda} \Delta \lambda
+ O\left((\Delta \lambda)^2\right)  
$
into Eq.~(\ref{D10}) and using 
$  {\rm Tr}\left[\frac{\dd \hat{\rho}(\lambda)}{\dd \lambda}\right]=
\frac{\dd }{\dd \lambda}{\rm Tr}\left[\hat{\rho}(\lambda)\right]=0$ 
and 
$\ln \hat{\rho}(\lambda+\Delta\lambda)  = \ln\hat\rho(\lambda)
+ \hat\rho^{-1}(\lambda)\frac{\dd \hat{\rho}(\lambda)}{\dd \lambda}\Delta\lambda+O\left( (\Delta \lambda)^2 \right)$.  
In the latter, the form of the density matrix operator
is assumed as in Eq.~(\ref{can}), and thus 
$[\hat{\rho}(\lambda),\frac{\dd \hat{\rho}(\lambda)}{\dd \lambda}]=0$. 
Since $D_{\lambda}(\Delta \lambda)>0$ for $\Delta \lambda \not = 0$, 
the vanishing of the first derivative in Eq.~(\ref{D_1vanish}) 
implies that $D_{\lambda}(\Delta \lambda)$
is differentiable at $\Delta \lambda=0$  (see Fig.~\ref{Dlambda}),
as discussed in Ref.~\cite{Blanco2013}. 
We thus find that 
\begin{equation}
  D_{\lambda}(\Delta \lambda)
  =
  O\left( (\Delta \lambda)^2 \right).
  \label{eq.Dvanish}
\end{equation}

\section{Effective temperature in subsystem $B$}~\label{app.TB}
In Sec.~\ref{sec.results}, we have found excellent agreement between 
statistical-mechanical quantities such as 
the thermodynamic entropy $S_A(\beta)$ and the internal energy $E_A(\beta)$ for an isolated subsystem $A$ 
and quantum-mechanical quantities such as 
the entanglement von Neumann entropy $\mathcal S_A(\lambda)$ and the energy $\mathcal{E}_A(\lambda)$
of a partitioned subsystem $A$ for a ground state of the whole system $A+B$, 
provided that the thermodynamic temperature $T=1/\beta$ in the former is set properly to 
the effective temperature $\mathcal{T}_{A}(\lambda)$ determined in the latter. 
A natural question is now 
how the effective temperature $\mathcal{T}_{B}(\lambda)$ of subsystem $B$ behaves.    
Here, $1/\mathcal{T}_{B}(\lambda)$ is defined as in Eq.~(\ref{beta_dSdE})
but with the energy $\mathcal{E}_B(\lambda)$ of subsystem $B$ instead of $\mathcal{E}_A(\lambda)$ of subsystem $A$ 
[note that $\mathcal{S}_B(\lambda) = \mathcal{S}_A(\lambda)$].
Because of the interaction term $\hat{V}_{AB}(\lambda)$, 
there exists quantum entanglement between subsystems $A$ and $B$.
This is different from the case discussed in Sec.~\ref{sec.assumption}, where no interactions are 
assumed between subsystems $X$ and $Y$, 
and hence $\mathcal{T}_A(\lambda) \not = \mathcal{T}_B(\lambda)$ is expected in general.

\begin{figure}
  \begin{center}
    \includegraphics[width=0.75\columnwidth]{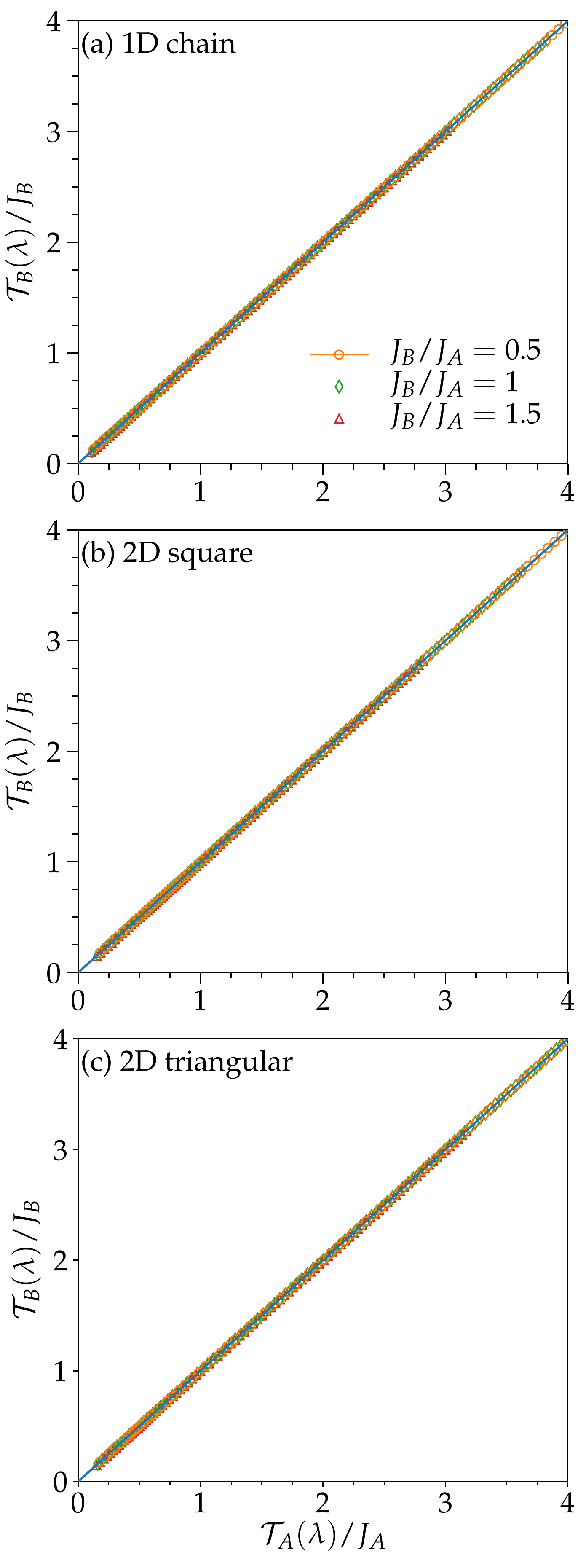}
    \caption{
      The effective temperature $\mathcal{T}_{B}(\lambda)/J_{B}$ of subsystem $B$
      as a function of
      the effective temperature $\mathcal{T}_{A}(\lambda)/J_{A}$ of subsystem $A$
      for the spin-1/2 antiferromagnetic Heisenberg model in 
      (a) the two coupled 1D chains (i.e, two-leg ladder), 
      (b) the two coupled 2D square lattice (i.e., bilayer square lattice), and 
      (c) the two coupled 2D triangular lattice (i.e., bilayer triangular lattice)  
      with $N_A=N_B=8$. 
      \label{TvsT}}    
  \end{center}
\end{figure}

Figure~\ref{TvsT} shows the effective temperature
$\mathcal{T}_{B}(\lambda)/J_{B}$ of subsystem $B$ as a function of
$\mathcal{T}_{A}(\lambda)/J_{A}$ for the three different lattices studied in Sec.~\ref{sec.results}. 
We find that   
$\mathcal{T}_{A}(\lambda)/J_{A} \simeq  \mathcal{T}_{B}(\lambda)/J_{B}$
even for $J_B/J_A\not=1$. 
Apparently, this relation is somewhat similar to
the common temperature condition in thermodynamics, 
which is a consequence of the equilibrium between two subsystems.
However, the important distinction is that here 
the microscopic energy scales $J_{A}$ and $J_{B}$ of subsystems $A$ and $B$, respectively, 
which are absent in thermodynamics,  
enter in the relation. 
Moreover, the relatively simple relation between
$\mathcal{T}_{A}(\lambda)$ and
$\mathcal{T}_{B}(\lambda)$ found here might be due to 
our setting of the Hamiltonian where $\hat{H}_{B} = (J_B/J_A) \hat{H}_{A}$. 
Finally, we note that the same relation is exactly satisfied in the two analytical 
examples discussed in Sec.~\ref{sec:analytical} 
[see Eqs.~(\ref{eq:temp_B}) and (\ref{eq:temp_F})].

\section{Thermofield-double-like state}\label{app.TFD}
In terms of the Schmidt decomposition
of the ground-state wavefunction $|\Psi_0(\lambda)\rangle$, 
the assumption in Eq.~(\ref{can})
along with Eq.~(\ref{eq:beta_a}) can be rephrased as 
\begin{equation}
  |\Psi_0(\lambda)\rangle
  \overset{!}{=}
  \sum_{n=1}^{D_A}
  \frac{
    \e^{-\mathcal{B}_A(\lambda) \epsilon^A_n/2}}
       {\sqrt{Z_A(\mathcal{B}_A)}}
       |\psi^A_n\rangle_A
       |g_n\rangle_B,
  \label{Schmidt}
\end{equation}
where
$|\psi^A_n\rangle_A$ is the $n$th eigenstate of $\hat{H}_A$ with its eigenvalue $\epsilon_n^A$ and 
$\{|g_n\rangle_B\}$ is a orthonormal basis set of subsystem $B$. 
The purification of $\hat{\rho}_{A}^{\rm red} (\lambda)$  
in Eq.~(\ref{Schmidt}) resembles to the thermofield double state~\cite{Fano1957}.
Indeed, if the subsystem $B$ is selected to be identical
to the subsystem $A$, the right hand side in Eq.~(\ref{Schmidt})
should reproduce the thermofield double state for the subsystem $A$ 
at temperature $1/\mathcal{B}_A(\lambda)$.

\bibliography{biball}

\end{document}